\documentclass[%
 reprint,
superscriptaddress,
 amsmath,amssymb,
 aps,
 prx,
]{revtex4-2}

\usepackage{placeins}
\usepackage{graphicx}
\usepackage{dcolumn}
\usepackage{bm}
\usepackage[colorlinks=true,allcolors=blue]{hyperref}
\usepackage{xcolor}

\newcommand{\dd}{\mathrm{d}}
\newcommand{\br}{\boldsymbol{r}}
\newcommand{\bcdot}{\boldsymbol{\cdot}}
\newcommand{\bx}{\boldsymbol{x}}
\newcommand{\bZ}{\boldsymbol{Z}}

\newcommand{\bu}{\boldsymbol{u}}
\newcommand{\bP}{\boldsymbol{P}}
\newcommand{\bB}{\boldsymbol{B}}
\newcommand{\bb}{\boldsymbol{b}}
\newcommand{\bA}{\boldsymbol{A}}
\newcommand{\bpsi}{\boldsymbol{\psi}}
\newcommand{\bJ}{\boldsymbol{J}}
\newcommand{\bk}{\boldsymbol{k}}
\newcommand{\bL}{\boldsymbol{L}}
\newcommand{\bnabla}{\boldsymbol{\nabla}}
\newcommand{\p}{\partial}
\newcommand{\mcE}{\mathcal{E}}

\newcommand{\const}{\mathrm{const}}
\newcommand{\Alfven}{Alfv\'{e}n}
\newcommand{\Elsasser}{Els\"{a}sser}

\begin{document}

\preprint{APS/123-QED}

\title{Reconnection-controlled decay of magnetohydrodynamic turbulence \\and the role of invariants}

\author{David N. Hosking}
 \email{david.hosking@physics.ox.ac.uk}
 \affiliation{Oxford Astrophysics, Denys Wilkinson Building, Keble Road, Oxford OX1 3RH, UK}
 \affiliation{Merton College, Merton Street, Oxford, OX1 4JD, UK
}
\author{Alexander A. Schekochihin}%

\affiliation{Merton College, Merton Street, Oxford, OX1 4JD, UK
}
\affiliation{The Rudolf Peierls Centre for Theoretical Physics, University of Oxford, Clarendon Laboratory, Parks Road, Oxford, OX1 3PU, UK}%

\date{\today}

\begin{abstract}
We present a new theoretical picture of magnetically dominated, decaying turbulence in the absence of a mean magnetic field. With direct numerical simulations, we demonstrate that the rate of turbulent decay is governed by the reconnection of magnetic structures, and not necessarily by ideal dynamics, as has previously been assumed. We obtain predictions for the magnetic-energy decay laws by proposing that turbulence decays on reconnection timescales, while respecting the conservation of certain integral invariants representing topological constraints satisfied by the reconnecting magnetic field. As is well known, the magnetic helicity is such an invariant for initially helical field configurations, but does not constrain non-helical decay, where the volume-averaged magnetic-helicity density vanishes. For such a decay, we propose a new integral invariant, analogous to the  Loitsyansky and Saffman invariants of hydrodynamic turbulence, that expresses the conservation of the random (scaling as $\mathrm{volume}^{1/2}$) magnetic helicity contained in any sufficiently large volume. We verify that this invariant is indeed well-conserved in our numerical simulations. Our treatment leads to novel predictions for the magnetic-energy decay laws: in particular, while we expect the canonical $t^{-2/3}$ power law for helical turbulence when reconnection is fast (i.e., plasmoid-dominated or stochastic), we find a shallower $t^{-4/7}$ decay in the slow `Sweet-Parker' reconnection regime, in better agreement with existing numerical simulations. For non-helical fields, for which there currently exists no definitive theory, we predict power laws of $t^{-10/9}$ and $t^{-20/17}$ in the fast- and slow-reconnection regimes, respectively. We formulate a general principle of decay of turbulent systems subject to conservation of Saffman-like invariants, and propose how it may be applied to MHD turbulence with a strong mean magnetic field and to isotropic MHD turbulence with initial equipartition between the magnetic and kinetic energies.
\end{abstract}

\maketitle


\section{\label{sec:intro}Introduction}


The nature of the decay of magnetohydrodynamic (MHD) turbulence is an important outstanding problem in fluid dynamics, with far-reaching consequences in astrophysics, from the evolution of primordial magnetic fields in cosmology~\cite{BanerjeeJedamzik04, DurrerNeronov13, Subramanian16} to the dynamics of the solar wind \cite{Chen11}. Naturally, the subject of decaying turbulence is one with a long history. In the hydrodynamic case, the basic problem of determining the exponent of the energy-decay power law was solved by Kolmogorov, in the third of his seminal 1941 papers on turbulence~\citep{Kolmogorov41c}. Kolmogorov's approach can be summarised as follows: (\textit{i}) identify an ideal invariant that is better conserved than the kinetic energy, then (\textit{ii}) posit a decay of the kinetic energy, occurring on the dynamical timescale, that conserves that invariant. In the case of hydrodynamic turbulence, Kolmogorov identified the Loitsyansky integral, 
\begin{equation}
    I_{\bL} = -\int \dd^3 \br \,r^2 \langle \bu(\bx) \bcdot \bu(\bx+\br)\rangle, \label{Loitsyansky}
\end{equation} as the relevant invariant. Physically, the conservation of $I_{\bL}$ represents the net conservation of angular momentum of the turbulent eddies~\cite{LandauLifshitzFluids, Davidson13}. Note, though, that the invariant controlling the decay is not simply the mean angular momentum, $\langle\bL\rangle= \langle \br \times \bu \rangle$, which, while conserved, is zero by isotropy. The Loitsyansky integral is therefore an example of an invariant that encodes the conservation of a quantity that individual eddies are expected to possess, but that has vanishing mean value due to its randomly directed nature.

Eq.~\eqref{Loitsyansky} implies the scaling $U^2 L^5 \sim \const$, where $L$ is the correlation scale of the turbulence, and $U$ is the typical velocity at that scale. Together with the identification of the dynamical timescale as $\tau \sim L/U$, this is enough to fix the decay rate of the kinetic energy $E= U^2/2$, as
\begin{equation}
    \frac{\dd E}{\dd t} \sim -\frac{E}{\tau}\sim  -\frac{E^{3/2}}{L}\sim -\, E^{17/10}, \label{Kolmogorov_decay}
\end{equation}which results in Kolmogorov's famous decay law 
\begin{equation}
    E\sim t^{-10/7}. \label{t10/7}
\end{equation}This result has been confirmed to excellent precision numerically~\cite{Ishida06, Davidson13}.



In this paper, we show how Kolmogorov's philosophy can be adapted to MHD turbulence. Despite extensive studies of the self-similar decaying solutions admitted by the MHD equations \cite{Hossain95, Matthaeus96, Wan12}, two problems have hindered past attempts to predict specific decay laws. First, there appear to exist a number of different regimes, depending on properties of the initial conditions \cite{StriblingMatthaeus90, StriblingMatthaeus91, Ting86, Wan12}. Restricting attention, for the moment, to the magnetically dominated case, where magnetic energy is much greater than kinetic, there are two canonical possibilities. First, there are helical field configurations, where the volume-averaged magnetic-helicity density, $\langle h \rangle = \langle \bA \bcdot \bB \rangle$, is non-zero. Then, conservation of $\langle h \rangle$ (a phenomenon sometimes referred to as `selective decay' \cite{Taylor74, Montgomery78, MatthaeusMontgomery80, Riyopoulos82, MatthaeusMontgomery84, Taylor86, Ting86}) provides a scaling that can be used to constrain the decay laws, $B^2 L \sim \const$ \cite{Hatori84, BiskampMuller99, Son99}, where $B$ is the typical size of the magnetic field at the correlation scale $L$. However, magnetic helicity is not sign-definite, so there also exist non-helical field configurations, for which $\langle h\rangle \ll B^2 L$. For such fields, the conservation of $\langle h\rangle$ does not impose a constraint on the decay. In this work, we show that even in such cases, the decay is still controlled by helicity conservation, in a manner formally analogous to the control of the decay of hydrodynamic turbulence by angular-momentum conservation.


Second, since, besides velocity, MHD has an additional field, $\bB$, there is no longer a dimensional inevitability in the identification of the decay timescale, as there was in hydrodynamics. Previous treatments~\cite{Hatori84, BiskampWelter89, BiskampMuller99, Son99} have assumed the Alfv\'{e}nic scaling $U\sim B$ in order to determine the ideal timescale uniquely \footnote{A different prescription for determining the decay timescale has also been considered \cite{Galtier97, Galtier99}, based on the so-called Iroshnikov-Kraichnan (IK) phenomenology of MHD turbulence \cite{Iroshnikov63, Kraichnan65}. Its predictions, however, do not appear to agree with later numerical work \cite{BiskampMuller99, BiskampMuller00, Christensson01, BanerjeeJedamzik04, FrickStepanov10, BereraLinkmann14, Brandenburg17,ReppinBanerjee17, Brandenburg19, Bhat20}. For example, the model predicts very slow growth of the energy-containing scale, $L \propto t^{1/6}$, in contrast to the prediction based on helicity conservation, $L \propto t^{2/3}$ \cite{BiskampMuller99}. The IK-based model furthermore does not predict a difference between the decay laws resulting from helical and non-helical initial conditions.}, though this is not well reproduced in numerics, where $B\gg U$ appears to be maintained if it was true initially, and, furthermore, for helical magnetic fields, a faster decay of the kinetic energy than the magnetic energy is often observed~\cite{BiskampMuller99,BiskampMuller00,Christensson01, BanerjeeJedamzik04, FrickStepanov10, BereraLinkmann14, Brandenburg17}. 

In fact, it is intuitively clear that relaxation on ideal timescales may not be possible for a strong initial magnetic field, because of the topological constraints imposed by magnetic-flux freezing. As was hypothesised by J.B.~Taylor, magnetic fields with non-trivial topologies relax via the reconnection of magnetic field lines \cite{Taylor74, Taylor86}, which transfers magnetic energy to larger scales. Reconnection, therefore, provides a physical explanation for the inverse transfer of magnetic energy observed in both helical and non-helical decaying MHD turbulence~\cite{BiskampBremer94, BiskampMuller00, Muller12, Zrake14, Brandenburg15, ReppinBanerjee17, Park17, Zhou20, Bhat20}. However, unless the values of the dissipation coefficients are sufficiently small for reconnection to occur in the plasmoid-dominated~\cite{Uzdensky10} or stochastic~\cite{Lazarian20} regime, magnetic reconnection occurs in the so-called `Sweet-Parker' regime~\cite{Sweet58,Parker57}, and is \emph{slow}, i.e., has a rate that is proportional to a negative fractional power of the Lundquist number, $S=BL/\eta$, where $\eta$ is the fluid resistivity. It should then be expected that the decay will proceed on the Sweet-Parker reconnection timescale, not the ideal one. The critical Lundquist number at which magnetic reconnection becomes fast (i.e., independent of $S$) is very large, $\sim 10^4$ \cite{Uzdensky10}, much larger than the typical Reynolds numbers at which Eq.~\eqref{t10/7} becomes a good description of the decay of hydrodynamic turbulence ($\mathrm{Re}\simeq 10^2$ \cite{Ishida06}). The requirement of such a large Lundquist number (corresponding to very thin current sheets), together with the large scale separation between the box size and the energy-containing scale needed to eliminate finite-box-size effects, mean that direct numerical simulations aiming to measure decay laws will generally be in the slow-reconnection regime \footnote{In two dimensions, the computational cost is becoming just about affordable --- see the study of MHD turbulence in the plasmoid-dominated regime by \cite{Dong18}. Even there, however, only a moderate scale separation between the box and energy-containing scales was possible, so it is doubtful such simulations could give a measurement of the decay rate free of finite-box-size effects.}. In~\cite{Zhou19}, it was demonstrated that two-dimensional MHD turbulence indeed decays on the Sweet-Parker timescale, and similar evidence has been presented for three-dimensional, non-helical turbulence too~\cite{Bhat20}, though it was interpreted as arising from the  two-dimensional decay mechanism put forward by~\cite{Zhou19}. One of the main goals of the present work will be to verify the reconnection-controlled nature of the three-dimensional decay in both the helical and non-helical cases, and to establish the corresponding decay laws for both energies.

Sweet-Parker reconnection is defined by the conditions of (\textit{i}) efficient conversion of magnetic energy to kinetic energy of reconnection outflows, and (\textit{ii}) a balance between the inductive term, $\nabla \times (\bu \times \bB)$, and the resistive dissipation term in the MHD induction equation, so that reconnection occurs in a time-invariant manner. This last requirement means that reconnection-controlled decaying turbulence in the Sweet-Parker regime is very different from decaying hydrodynamic turbulence, in that it is sensitive to the precise form of the dissipation term. Importantly, this means that different decay power laws are expected in numerical simulations depending on whether Laplacian dissipation, $ \propto \eta\nabla^2 \bB \equiv \eta_2\nabla^2 \bB$, or hyper-dissipation, $\propto \eta_n\nabla^n \bB$, is employed. This fact has not been widely appreciated, and leads to a simple test of whether reconnection indeed governs the decay timescale: simulations at moderately large (but not so large so as to be in the `fast' reconnection regime) Lundquist numbers should exhibit different decay power laws depending on the order of hyper-dissipation, $n$. In this paper, we present such simulations, conducted with the incompressible, spectral MHD code Snoopy~\cite{Lesur15}. They turn out to be in excellent agreement with these expectations. 

An outline of the rest of this paper is as follows. In Section~\ref{sec:helical}, we consider the decay of helical MHD turbulence from a magnetically dominated state. Since magnetic helicity is better conserved than magnetic energy in the limit of vanishing resistivity, helicity is precisely a quantity that can control the evolution in the same way as the Loitsyansky integral does in hydrodynamics. Applying a Kolmogorov-style argument to helicity conservation for a decay occuring on the ideal timescale, $L/B$, yields a power law of $t^{-2/3}$ for both the magnetic and kinetic energies~\cite{Hatori84, BiskampMuller99, Son99}, though this decay law is not well-supported by numerics, where a shallower decay law for magnetic energy, and a steeper decay law for kinetic energy are typically observed~\cite{BiskampMuller99, BiskampMuller00, Christensson01, BanerjeeJedamzik04, FrickStepanov10, BereraLinkmann14, Brandenburg17, Brandenburg19}. We show that both of these results are expected for a decay occurring via magnetic reconnection in the Sweet-Parker regime. In particular, we show that the magnetic energy should decay as $t^{-4/7}$, with $t^{-2/3}$ only achieved by fast reconnection. We also find that, provided the dominant flows are contained within Sweet-Parker sheets, the faster decay of kinetic energy, as $t^{-5/7}$, is a natural consequence of their changing aspect ratio.

In Section~\ref{sec:non-helical}, we consider the decay of non-helical MHD turbulence from a magnetically dominated state. The mechanism controlling this type of decay has so far remained unknown: because the mean helicity density vanishes, its conservation cannot be used to derive a scaling relation relating $B$, the characteristic magnetic field, to its characteristic length scale $L$. Numerically, decay laws for both the magnetic and kinetic energies of close to $t^{-1}$ have been observed~\cite{Zrake14, Brandenburg15, Brandenburg17, ReppinBanerjee17,Bhat20}, though there is no definitive theoretical explanation for this behaviour. An influential idea is that in the absence of an integral invariant, the decay might satisfy the well-known scaling symmetry of the MHD equations~\cite{Olesen97}, including dissipative terms. Such a decay would have a $t^{-1}$ power law for the magnetic energy~\cite{Orszag77, Hossain95, Matthaeus96, Campanelli04}. Another suggestion is that the non-helical decay is effectively two-dimensional. In this case, it is the conservation of `anastrophy', or the square of the magnetic vector potential, that should control the decay~\cite{MatthaeusMontgomery80, Ting86}. A Kolmogorov-style argument then leads to a $t^{-1}$ power law independently of whether the decay occurs on the ideal~\cite{Hatori84, BiskampWelter89, Brandenburg15, Brandenburg17}, or the Sweet-Parker~\cite{Zhou19, Zhou20, Bhat20} timescale (see Appendix~\ref{sec:2d}). That both treatments predict the same power law is a coincidence related to the fact that anastrophy conservation implies constant Lundquist number for $n=2$ resistive dissipation \cite{Zhou19} (incidentally, the scaling argument is essentially this same point with the direction of implication reversed). However, it is not clear why fully three-dimensional, isotropic turbulence should two-dimensionalise in this way, nor why any special significance should be given to constant Lundquist number. Indeed, we show in Section~\ref{sec:non-helical} that both are inconsistent with numerical evidence.

Instead, we propose a treatment of the non-helical decay controlled by the conservation of fluctuations in magnetic helicity. The key point is that the vanishing of the total magnetic helicity does not necessarily imply that any given magnetic field structure is non-helical. Indeed, non-helical magnetic structures generally relax on ideal timescales to zero magnetic energy, assuming they are not constrained by higher-order topological invariants. We therefore expect that the natural state of the turbulence will be to contain a collection of helical structures, though there will be equal abundances of positive- and negative-helicity structures so that the zero-overall-helicity constraint is satisfied. For such a turbulence, we identify a new integral invariant whose relation to magnetic helicity is precisely analogous to that of the Loitsyansky integral to angular momentum. We refer to this invariant as the `Saffman helicity invariant', due to the close analogy between it and the integral invariant proposed by Saffman for hydrodynamic turbulence~\cite{Saffman67}, and also between our arguments and the arguments usually associated with the Saffman invariant and its various anisotropic generalisations \cite{Davidson12, Davidson13}. As we will show, the conservation of our new invariant implies the scaling  $B^4 L^5 \sim\const$. This scaling implies a magnetic-energy-decay power law of $t^{-20/17}$ if reconnection occurs on the Sweet-Parker timescale, or $t^{-10/9}$ if reconnection is fast. These power laws are different from~$t^{-1}$, but are still in excellent agreement with published numerical results, and with our own numerical results presented below. We also find that, for non-helical magnetic fields, the rate at which the aspect ratio of the current sheets changes is much smaller than in the helical case, explaining why a faster decay of kinetic energy is not observed for a non-helical decay from initial states that have small kinetic energy.


Finally, in Section~\ref{sec:discussion}, we discuss the behaviour of systems with fractional helicity, which we show will ultimately always transition to the fully helical regime as long as the system size is sufficiently large. We also discuss possible applications of the Saffman formalism to wider classes of turbulent decays. As an example, we suggest the existence of a Saffman-type cross-helicity invariant that may control the critically balanced decay of MHD turbulence in the presence of a mean magnetic field, recently studied in \cite{Zhou20}. We also conjecture that the simultaneous conservation of both the Saffman-type cross-helicity invariant and the magnetic helicity might govern the initial period of decay of an MHD state starting with $U\sim B$.

\section{\label{sec:helical}Decay of helical turbulence}\FloatBarrier

We first consider the decay of MHD turbulence from an initial state where magnetic energy dominates kinetic ($B\gg U$), and the magnetic field is helical, i.e., the volume-averaged magnetic-helicity density,
\begin{equation}
    \langle h \rangle = \lim_{V \to \infty} \frac{1}{V}\,H_V = \lim_{V \to \infty} \frac{1}{V}\int_V \dd^3 \br \, \bA \bcdot \bB, \label{volume-average-helicity}
\end{equation}is large, $\sim B^2 L$. The case of initial parity between the two energies ($U\sim B$) will be discussed in Section~\ref{sec:UsimB}. Magnetic helicity is a topological invariant --- the total magnetic helicity of a collection of flux tubes is equal to the amount of (signed) flux linked by these tubes. As such, its conservation is related to \Alfven's theorem, which states that for $\eta_n = 0$, the magnetic field is frozen into fluid motions, ensuring that all topological invariants are precisely conserved. The special significance of magnetic helicity is that, unlike other topological invariants, it remains approximately conserved (i.e., better conserved than energy) for small but non-vanishing $\eta_n$ \cite{Ruzmaikin94}. This statement is true independently of the reconnection regime. A proof, adapted from \cite{Berger84}, is as follows.

\subsection{Helicity is conserved by \\(hyper-resistive) reconnection \label{sec:helicity_conservation}}

In hyper-dissipative MHD, the evolution of the magnetic helicity in a closed volume whose surface is everywhere normal to $\bB$ satisfies
\begin{equation}
    \left|\frac{\dd H}{\dd t}\right|=  2\eta_n \left|  \int\mathrm{d}^3\boldsymbol{r}\, (\nabla^{n}\bA)\bcdot \bB \right|. \label{helicity_evolution}
\end{equation}Note that setting $n=2$ here yields the familiar expression in terms of the current helicity. After integrating the integral on the right-hand side by parts, applying the Cauchy-Schwarz inequality, and then integrating by parts again, one obtains
\begin{equation}
    \left|\frac{\dd H}{\dd t}\right|^2 \leq  4\left|\frac{\dd E_M}{\dd t} \,\eta_n \int\mathrm{d}^3\boldsymbol{r}\,\bA \bcdot (\nabla^n \bA)\right|,\label{CauchySchwarz}
\end{equation} where $\dd E_M/\dd t = \eta_n\int\mathrm{d}^3\boldsymbol{r}\,\bB \bcdot (\nabla^n \bB)$ is the rate of magnetic-energy decay due to Ohmic heating. For $n=2$, the other integral in Eq.~\eqref{CauchySchwarz} is just twice the magnetic energy~\cite{Berger84}. More generally, we can write
\begin{align}
    \eta_n\left|\int\mathrm{d}^3\boldsymbol{r}\,\bA \bcdot (\nabla^n \bA)\right| &= \eta_n\left|\int\mathrm{d}^3\boldsymbol{r}\,\bB \bcdot (\nabla^{n-2} \bB)\right|\nonumber \\ &\sim \frac{\dd E_M}{\dd t} \,{\delta_\eta}^2,
\end{align} where $\delta_\eta$ is the resistive dissipation scale. Eq.~\eqref{CauchySchwarz} then implies
\begin{equation}
    \frac{\dd \log H}{\dd t} \sim  \frac{\delta_{\eta}}{L} \frac{\dd \log E_M}{\dd t}. \label{scaling_for_helicity_decay}
\end{equation}Eq.~\eqref{scaling_for_helicity_decay} states that the rate of change of magnetic helicity is smaller than the rate of the energy decay due to Ohmic heating (which will be even smaller than the true magnetic-energy-decay rate, because magnetic energy can also be converted to the kinetic energy of reconnection outflows) by a factor equal to the ratio of the integral scale to the resistive dissipation length scale, which becomes arbitrarily small as $\eta_n \rightarrow 0^+$. \textit{Q.E.D.}

It may appear counter-intuitive that reconnection, a process that, by definition, changes the topology of magnetic field lines, can conserve helicity, a topological invariant. The resolution of this apparent paradox is that self-linkages, i.e., twists of the magnetic flux tube are also associated with helicity. For example, during the unlinking of two linked tori by reconnection to form a single torus, the resulting torus ends up twisted, and the total helicity of the configuration is conserved \cite{Ruzmaikin94}.

\subsection{Theory of helical decay}


The conservation of the volume-averaged magnetic helicity, Eq.~\eqref{volume-average-helicity}, implies the scaling 
\begin{equation}
    B^2 L\sim \const.\label{B2L}
\end{equation}The remaining ingredient required to compute the magnetic-energy decay law is the decay timescale, as a function of $U$, $B$ and $L$. The simplest possible treatment is to assume Alfv\'{e}nic  dynamics, with $U\sim B$, similarity between the integral scales of the magnetic and kinetic energies, $L$, and, therefore, a decay timescale~$L/B\sim L/U$. For future reference, we compute the expected decay power law under such an assumption for a scaling more general than Eq.~\eqref{B2L}, \textit{viz.},
\begin{equation}
    B^\alpha L\sim \const. \label{BalphaL}
\end{equation}Eq.~\eqref{Kolmogorov_decay} becomes
\begin{equation}
    \frac{\mathrm{d}}{\mathrm{d} t} \frac{1}{2} B^2 \sim  -\frac{B^{3}}{L} \propto -B^{3+\alpha}, \label{ideal_DE}
\end{equation}with solution
\begin{equation}
    B^2 \sim t^{-2/(1+\alpha)}. \label{ideal_law}
\end{equation}Now setting $\alpha = 2$ for helicity conservation, the canonical $t^{-2/3}$ power law is recovered~\cite{Hatori84}. However, this prediction has proved to be in poor agreement with numerics, which have found a decay of the magnetic energy closer to $t^{-1/2}$, and, unexpectedly, a decay of the kinetic energy faster than $t^{-2/3}$~\cite{BiskampMuller99, BiskampMuller00, Christensson01, BanerjeeJedamzik04, FrickStepanov10, BereraLinkmann14, Brandenburg17, Brandenburg19}.

These discrepancies are readily resolved by assuming the decay to occur on the reconnection, rather than ideal, timescale. Naturally, the reconnection timescale depends on the reconnection regime, i.e., on whether the reconnection is slow (Sweet-Parker) or fast (plasmoid-dominated \cite{Uzdensky10} or stochastic \cite{Lazarian20}). Fast reconnection, by definition, occurs on dynamical timescales, so will again produce a $t^{-2/3}$ decay. However, as explained in the Introduction, extant numerical simulations of decaying MHD turbulence mostly probe the slow regime, owing to the large Lundquist numbers and hence large resolutions required for fast reconnection to take place. In this case, the reconnection timescale is
\begin{equation}
    \tau_\mathrm{rec} \sim  S_n^{1/n}\, \frac{L}{B}, \label{rectimescale}
\end{equation}where $S_n=BL^{n-1}/\eta_n$ is the hyper-Lundquist number. Using Eq.~\eqref{rectimescale} for the decay timescale, Eq.~\eqref{ideal_DE} becomes
\begin{equation}
    \frac{\mathrm{d}}{\mathrm{d} t} \frac{1}{2} B^2 \sim  -\frac{B^{3}}{L} B^{-1/n} L^{(1-n)/n} \propto -B^{(2\alpha n+3n-\alpha-1)/n}, \label{reconnecting_DE}
\end{equation}the solution of which is
\begin{equation}
    B^2 \sim t^{-2n/(2\alpha n + n - \alpha - 1)}. \label{reconnecting_plaw}
\end{equation}Again, for a helicity-conserving decay, we set $\alpha = 2$ to obtain a power law of 
\begin{equation}
   B^2 \sim t^{-2n/(5n-3)}. \label{generaln_helical}
\end{equation}For $n=2$ (Laplacian dissipation), the power law is
\begin{equation}
    E_M \sim t^{-p_M},\quad p_M = \frac{4}{7} \simeq 0.57,
\end{equation}which is indeed shallower than $p_M = 2/3$. Indeed, some recent studies at large resolution have reported $p_M \simeq 0.58$, in remarkable agreement with this prediction~\cite{Brandenburg17,Brandenburg19}. However, we caution against direct comparison with those simulations, because they employed time-dependent dissipation coefficients. For numerical studies using $n=4$ hyper-dissipation~\citep{BiskampMuller99}, Eq.~\eqref{reconnecting_plaw} predicts an even slower magnetic-energy-decay exponent of $p_M = 8/17 \simeq 0.47$. This is in excellent agreement with the $p_M \simeq 0.5$ found numerically in \cite{BiskampMuller99}.


It may appear counter-intuitive that reconnection can result in a faster decay of kinetic energy than magnetic energy, as reconnection outflows are typically Alfvénic, i.e., the outflow velocity is approximately equal to the upstream Alfv\'{e}n speed of the magnetic field prior to reconnection --- a condition that is hard-wired into the Sweet-Parker scalings. However, the current sheets, where reconnection occurs, are not volume-filling. Denoting the current sheet width by $\delta$, the volume occupied by the current sheet formed when two structures of volume $L^3$ reconnect and merge is $L^2 \delta$. Therefore, we expect 
\begin{equation}
    E_K \sim \frac{\delta}{L} E_M \label{EK_sim_dL_EM}
\end{equation}for Alfv\'{e}nic outflows, where $E_K$ and $E_M$ should be understood as the total kinetic and magnetic energies in the system, respectively. This result allows for the possibility of different decay rates for the kinetic and magnetic energy, because the ratio $\delta / L$ need not be constant in time. For example, with hyper-dissipative Sweet-Parker sheets, $\delta / L \sim S^{-1/n}_n\sim (BL^{n-1})^{-1/n}$. For $B^2 L \sim \const$ [Eq.~\eqref{B2L}], we find $\delta / L \sim E_M^{1-3/2n}$, which, via Eq.~\eqref{EK_sim_dL_EM}, translates to 
\begin{equation}
    E_K\sim E_M^{2-3/2n}. \label{kinetic_energy_decay}
\end{equation}Thus, kinetic energy is indeed expected to decay more quickly than magnetic energy, simply due to the changing aspect ratio of the Sweet-Parker sheets. For $n=2$, this effect is relatively modest: the kinetic-energy decay exponent is $5/4$ times greater than the magnetic one, so the decay exponents for $E_M$ and $E_K$ are $p_M=4/7\simeq 0.57$ and $p_K = 5/7\simeq  0.71$, respectively. For $n=4$, though, the kinetic-energy decay exponent is $13/8$ times greater than the magnetic one, so these decay exponents become $p_M=8/17\simeq 0.47$ and $p_K=13/17 \simeq 0.76$. 

We note that the validity of Eq.~\eqref{EK_sim_dL_EM} requires the initial kinetic energy to be much smaller than the magnetic energy, by a factor $\lesssim \delta/L$, as otherwise the kinetic energy contained in outflows may be subdominant to the rest. This condition will be true for our simulations, which are initialised with $U=0$, but may not be true in all situations of interest. However, as we shall discuss in Section~\ref{sec:UsimB}, there is reason to believe that decaying helical MHD turbulence should be driven towards Eq.~\eqref{EK_sim_dL_EM}, even if $E_K$ is larger than $(\delta/L) E_M$ initially.

\subsection{Numerical results\label{sec:helical_results}}



In this section, we compare the theoretical predictions of the previous section with numerical simulations of MHD turbulence, initialised as a Gaussian random magnetic field with characteristic scale $\simeq 1/33$ of the box size. These simulations employ the incompressible, spectral MHD code Snoopy~\cite{Lesur15}, with Prandtl number $\mathrm{Pm}\equiv\nu_n/\eta_n=1$. We provide further information about Snoopy and describe the details of our numerical setup in Appendix~\ref{sec:app_plaws}.

In simulations with $n=2$ dissipation, resolution constraints prevent the use of Lundquist numbers that are sufficiently large to achieve good conservation of the magnetic helicity. Nonetheless, supposing that helicity decays as $H(t)\sim t^{-p_H}$, while $E_M(t)\sim t^{-p_M}$, we expect $B^{2(1-p_H /p_M)}L\sim \const$. For small but non-vanishing $\eta_n$, therefore, we expect to find an $\alpha$ somewhat smaller than $2$ such that $B^\alpha L\sim \const$. We can determine this value of $\alpha$ numerically by measuring $E_M(t)$ and
\begin{equation}
    L = \frac{2\pi}{E_M} \int \dd k\, \frac{{\mcE}_M(k)}{k},
\end{equation}where ${\mcE}_M(k)$ is the spectral magnetic-energy density.

As long as $\eta_n$ is not too large, the decay should still occur on the Sweet-Parker timescale. We can then use Eq.~\eqref{reconnecting_plaw} to determine the expected magnetic-energy decay exponent based on the empirically determined value of $\alpha$, and compare it with the value measured in our simulations. While we do not expect that this procedure should yield exact agreement between the predicted and empirical decay exponents, as it neglects the role of Ohmic diffusion (which, when $S_n$ is small, will ultimately become more important to the decay than reconnection), we expect approximate agreement that becomes better as $S_n$ increases and $\alpha$, therefore, becomes closer to $2$.

\begin{figure}
\includegraphics[width=\columnwidth]{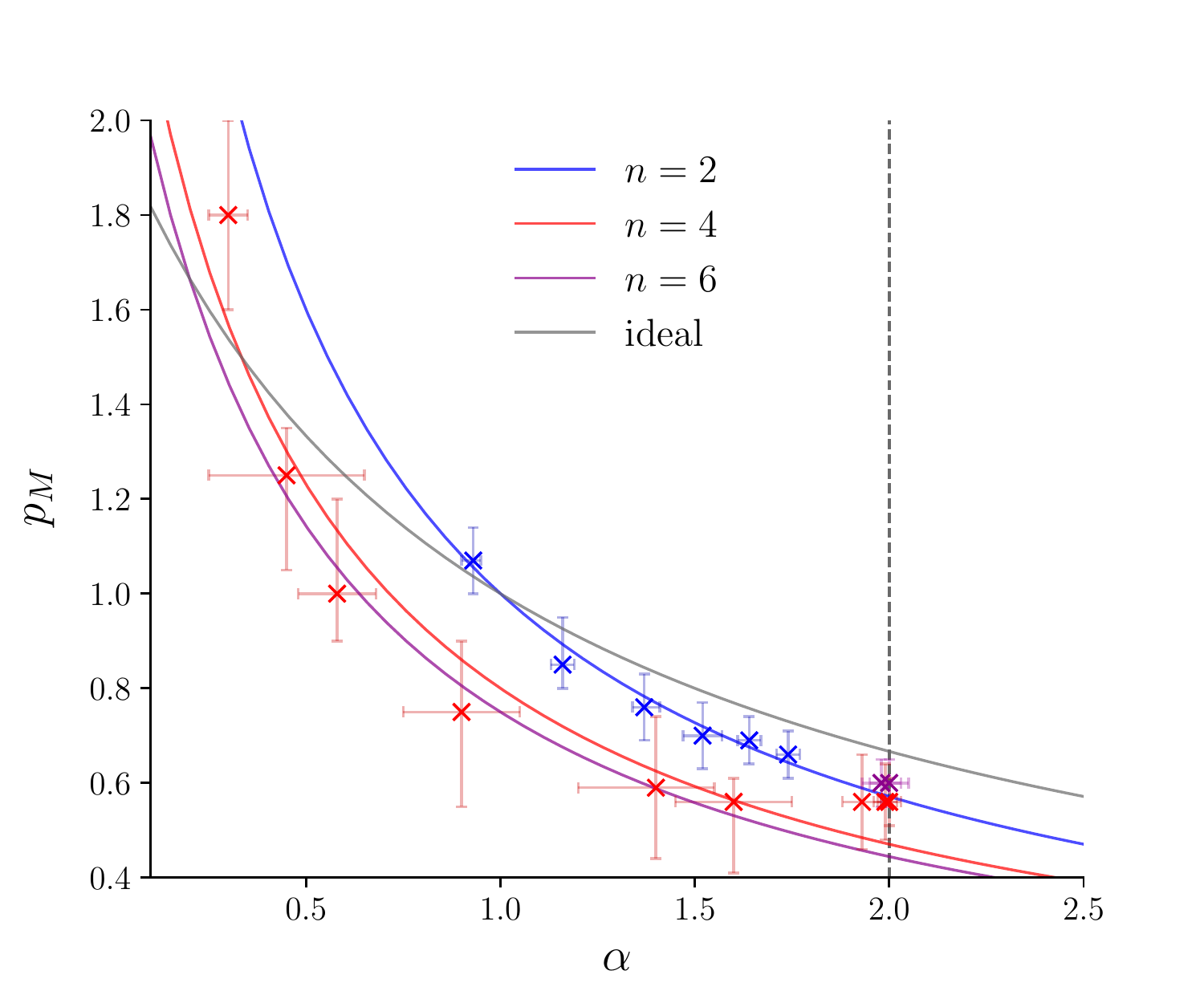}
\caption{\label{fig:p_alpha_helical} Relation between the empirically obtained magnetic-energy-decay power-law exponent, $p_M$, and the value of $\alpha$ for which $B^\alpha L \sim \const$. Solid curves show the expected relationship, Eq.~\eqref{reconnecting_plaw}, for decays occurring on the Sweet-Parker timescale, with $n=2$, $n=4$ and $n=6$ shown in blue, red, and magenta, respectively. The grey solid curve depicts the `ideal' scaling given by~\eqref{ideal_law}. Simulation results are in excellent agreement with the coloured curves, and not with the grey curve. This confirms that the decay takes place on the Sweet-Parker, rather than ideal, timescale. The full set of decay curves from which this plot was derived can be found in Appendix \ref{sec:app_plaws}. }
\end{figure}

\begin{figure}
\includegraphics[width=\columnwidth]{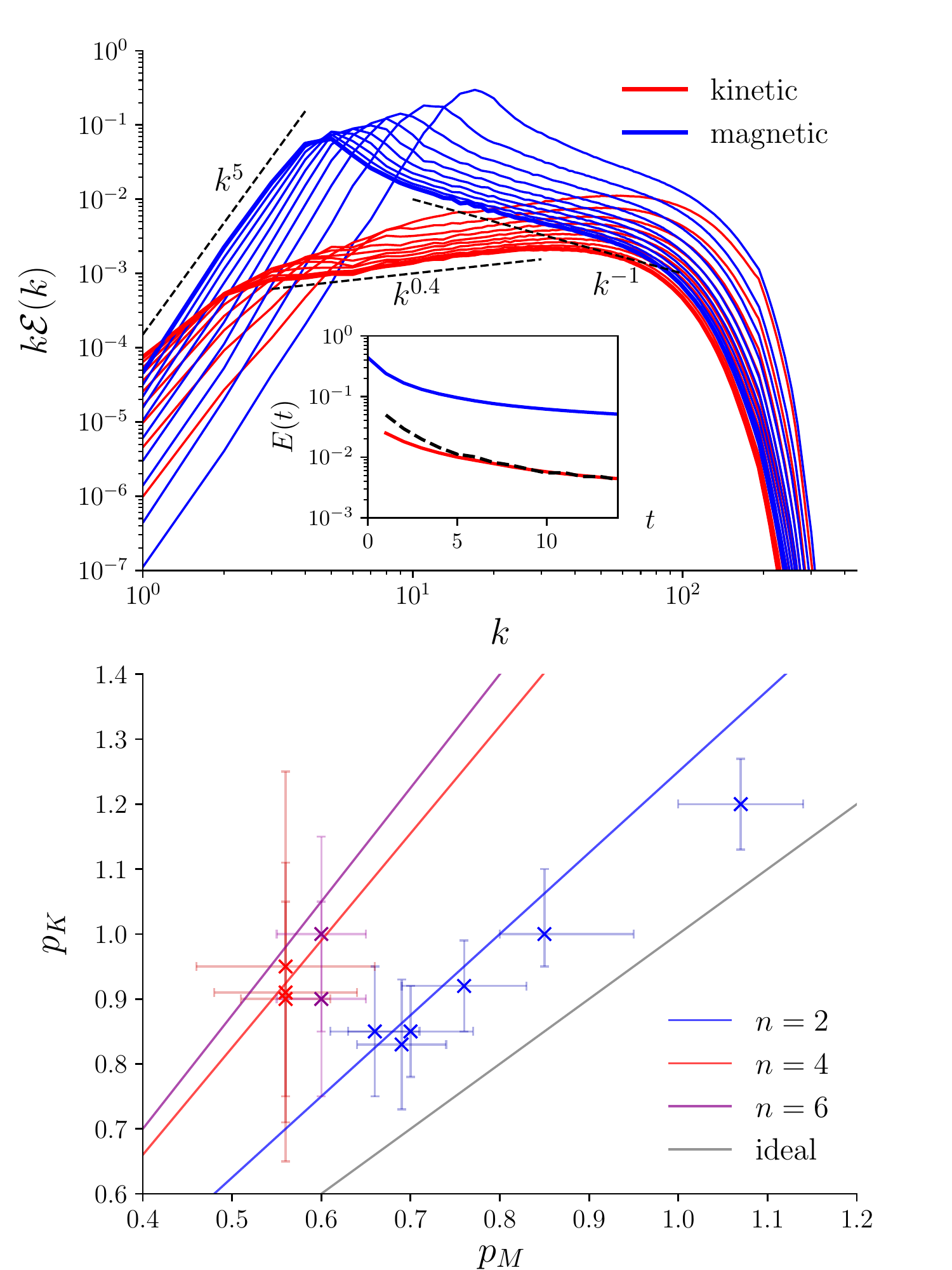}
\caption{\label{fig:helical_spectrum}Top panel: Evolution of $k\mcE(k)$, where $\mcE(k)$ is the spectral energy density, for the magnetic (blue) and kinetic (red) energies. These plots are obtained from a helical simulation with $n=4$, $\eta_4 = 2\times 10^{-8}$. Each plot of $k\mcE(k)$ is separated by a time interval of $1.0$ between $t=1.0$ to $t=10.0$ ($t=10.0$ in bold), where time is in code units based on normalising the box size and initial mean-square magnetic field to $2\pi$ and $1$, respectively, so that $1$ time unit is approximately the initial \Alfven~crossing time of the box. The peak of the magnetic energy is at much larger scales (smaller $k$) than the peak of the kinetic energy, consistent with the expectation that the kinetic energy should be contained within the Sweet-Parker sheets. Inset: decay of the total magnetic and kinetic energies. The total kinetic energy curve is much below the magnetic energy curve, and coincides with $(\delta/L)E_M$ (black), in agreement with Eq.~\eqref{EK_sim_dL_EM}, with $\delta/L$ computed as the ratio of the wavenumbers at which $k\mcE_M(k)$ and $k\mcE_K(k)$ peak.
Bottom panel: 
Plot of the kinetic-energy decay exponent, $p_M$, against the magnetic-energy decay exponent, $p_K$, as measured in simulations with $S_{n,\,0}^{1/n}>9.0$. Results are in reasonable agreement with the theoretical prediction, Eq.~\eqref{kinetic_energy_decay} (coloured lines), and are inconsistent with $E_K\propto E_M$ (grey line).}
\end{figure}

In Fig.~\ref{fig:p_alpha_helical}, we present the results of such a comparison, plotting the empirically measured values of $p_M$ against those of $\alpha$, for simulations with $n= 2$ and $n=4$ (how the error bars are determined is explained in Appendix~\ref{sec:app_plaws}). This figure shows remarkable agreement between our simulations and the Sweet-Parker decay curves (coloured), despite the fact that we do not reach the asymptotic value of $\alpha = 2$ with $n=2$ dissipation. For $n=4$, we also find excellent agreement, and do reach $\alpha = 2$. As other authors have noted~\cite{BiskampMuller99}, this asymptotic scaling is reached much more rapidly in the hyper-dissipative case, and indeed we were forced to choose relatively small values of $S_n^{1/n}$ in order to populate the part of Fig.~\ref{fig:p_alpha_helical} with $\alpha<2$. To illustrate this point further, we plot two simulations with $n=6$ hyper-dissipation, which also exhibit the $\alpha=2$ scaling (and are almost coincident in Fig.~\ref{fig:p_alpha_helical}). The faster attainment of the correct asymptotic scaling with hyper-dissipation will be important in establishing the correct value of $\alpha$ for non-helical turbulence, previously unknown, in the next section.

Interestingly, we note that for $\alpha \simeq 2$, the decay exponent $p_M$ is consistently somewhat larger than our theoretical prediction based on Sweet-Parker reconnection. This suggests these simulations may be at the start of the transition to the fast-reconnection regime. We do not find the same transition in two-dimensional simulations (see Appendix~\ref{sec:2d}), despite employing even larger Lundquist numbers, which is consistent with the intuitive expectation that fast reconnection should `turn on' more quickly in three dimensions, due to turbulence in the reconnection region.

\begin{figure*}
\includegraphics[width=\textwidth]{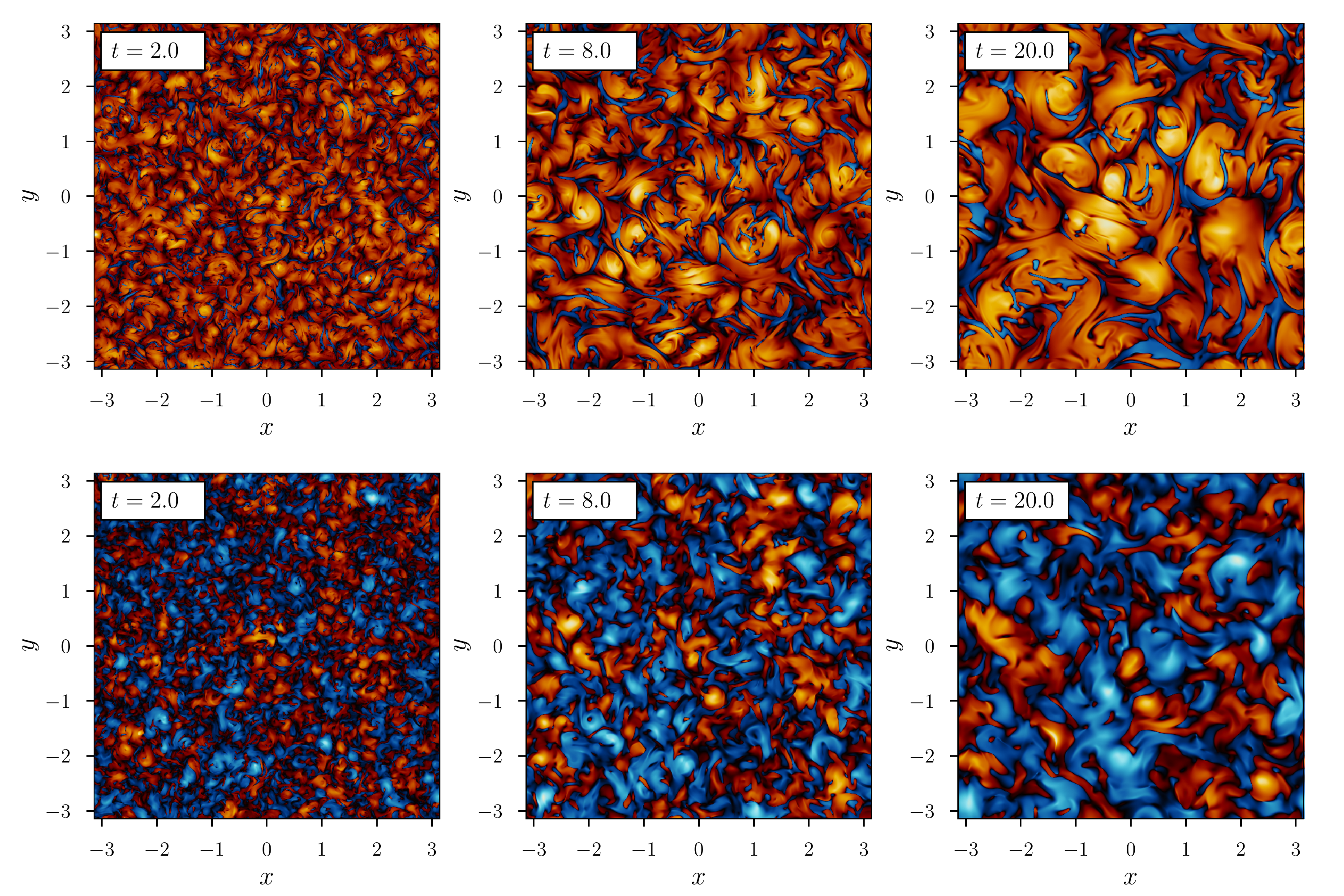}
\caption{\label{fig:3d_slices}Slices of $\bB \bcdot \hat{\bJ}$, the projection of the magnetic field onto the direction of the electric current ($\hat{\bJ}$ is the unit vector in this direction), for three different times during helical (above) and non-helical (below) simulations with $n=2$, $\eta_2 = 2\times 10^{-4}$ (time is in code units, as explained in the caption to Fig.~\ref{fig:helical_spectrum}). Positive values are shown in red, negative values in blue. $\bB \bcdot \hat{\bJ}$ is a local measure of the twist of the magnetic field lines, with different signs indicating different directions of the twist. It is also related to the magnetic helicity, because the sign of $\bB \bcdot \hat{\bJ}$ is equal to the sign of the magnetic helicity for a fully relaxed helical magnetic structure, according to the J.B.~Taylor relaxation theory~\cite{Taylor74, Taylor86}. Indeed, our helical simulations do feature a super-abundance of blobs with positive $\bB \bcdot \hat{\bJ}$, while in our non-helical simulations, blobs with both signs of $\bB \bcdot \hat{\bJ}$ have approximately equal representation.}
\end{figure*}

\begin{figure*}
\includegraphics[width=\textwidth]{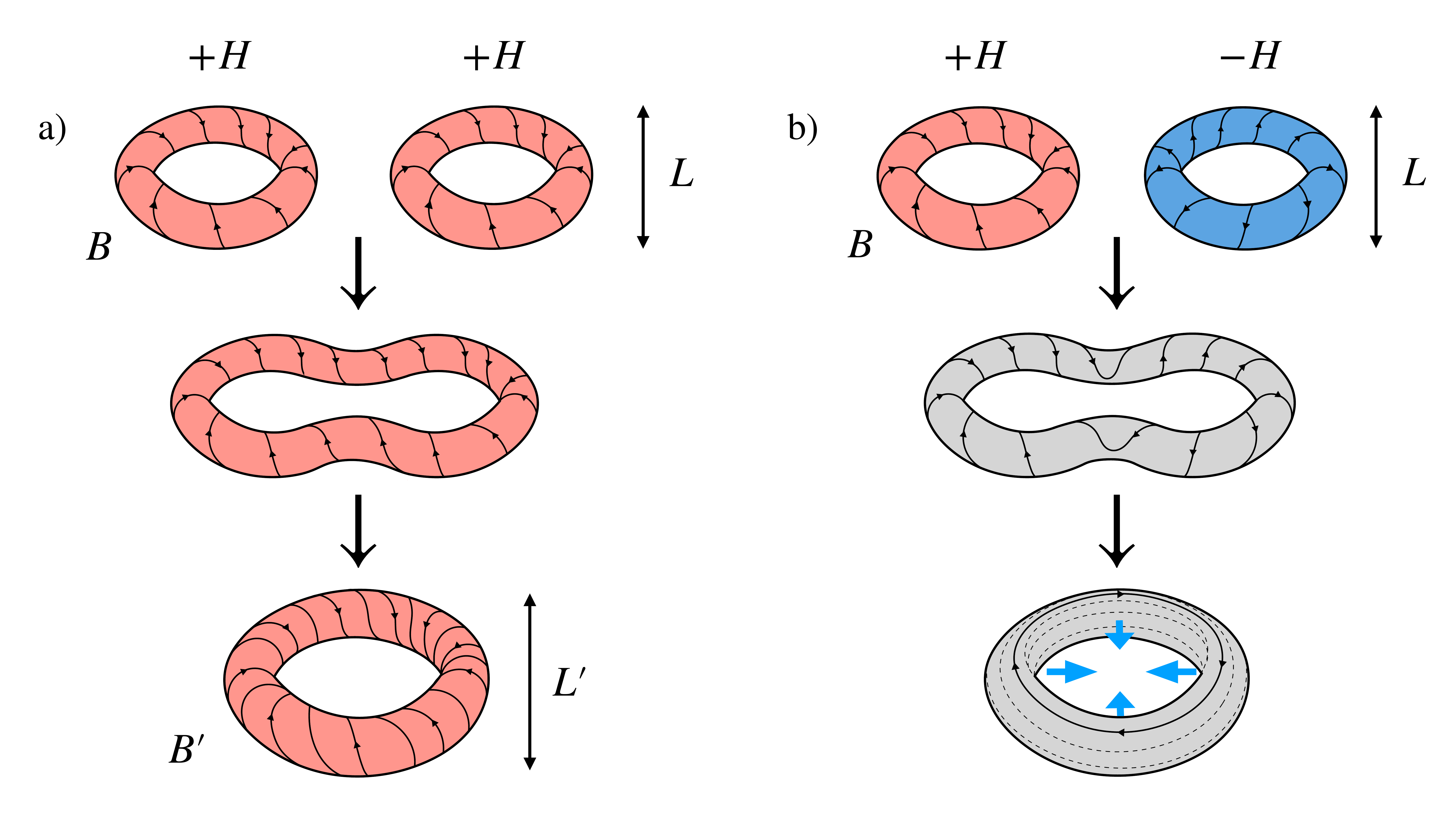}
\caption{\label{fig:cartoon} (a) Cartoon of a typical merger of two helical structures (`blobs'). As explained in the main text, $B'^2 L' = B^2 L$ for helical turbulence in which this is the only allowed process.
(b) This additional `annihilation' process is also possible in non-helical turbulence, and should occur equally frequently. The presence of this process modifies the previous scaling to $B'^4 L'^5 = B^{4} L^5$.}
\end{figure*}

For the decay of the kinetic energy also, we find that the predictions of the previous section are in good agreement with our simulations, as shown in Fig.~\ref{fig:helical_spectrum}. The upper panel shows the evolution of the magnetic- and kinetic-energy spectra for a run with $n=4$, $\eta_4 = 2\times 10^{-8}$, confirming that at any given time, kinetic energy is contained at much smaller scales than the magnetic energy, consistent with the expectation that reconnection outflows should have a width $\delta \ll L$. The inset shows the relative sizes of the total magnetic and kinetic energies in the same run, which are in excellent agreement with  Eq.~\eqref{EK_sim_dL_EM}. The lower panel shows the kinetic-energy decay exponents, $p_K$, plotted against the corresponding exponents for the magnetic energy, $p_M$, confirming that a faster decay of kinetic energy is realised in our simulations, in reasonably good agreement with our theoretical prediction, Eq.~\eqref{kinetic_energy_decay}.

We note that, while we do observe faster decay of kinetic energy than magnetic in our simulations, the difference is not as stark as in the numerical study by \cite{BiskampMuller99}, who found $E_K \propto E_M^2$. As we will discuss in Section \ref{sec:discussion}, this discrepancy may arise because the initial state in their study was one with equipartition between the magnetic and kinetic energies, $U\sim B$, unlike the $U\ll B$ we have employed here. We will argue that when $U\sim B$,  the conservation of cross-helicity, even though the latter is a sign-indefinite quantity, might play an important role in governing the decay. Simultaneously conserving magnetic helicity and \textit{fluctuations} in the cross-helicity, under the same formalism as we are about to propose for helicity conservation in non-helical turbulence, does imply the scaling $E_K \propto E_M^2$ conjectured in \cite{BiskampMuller99}.


To conclude, the results of this section represent what we consider compelling evidence that helical, magnetically dominated MHD turbulence relaxes by the self-similar coalescence of magnetic structures via magnetic reconnection, as was suggested by J.B.~Taylor \cite{Taylor86}. Physically, this implies that the correct way to think about the system is as consisting of a collection of magnetic structures that are unable to relax under ideal dynamics due to the topological constraints imposed by the flux freezing, and, therefore,  relax via coalescence on a timescale at which these constraints can be broken, i.e., on the reconnection timescale. 

\section{\label{sec:non-helical}Decay of non-helical turbulence}

Having established a theory of the reconnection-controlled decay of helical turbulence from a magnetically dominated state, we now consider the case of non-helical turbulence, i.e., turbulence for which the volume-averaged magnetic-helicity density [Eq.~\eqref{volume-average-helicity}] vanishes. Here, as in Section~\ref{sec:helical}, we consider the decay from an initial state with predominant magnetic energy ($B\gg U$), postponing the discussion of the case with $U\sim B$ until Section~\ref{sec:UsimB}. As we have already noted, the mechanisms controlling the evolution of such turbulence are not well understood. Numerically, a power law close to $t^{-1}$ has been measured~\cite{Zrake14, Brandenburg15, Brandenburg17,ReppinBanerjee17, Bhat20}, prompting comparisons with the two-dimensional decay~\cite{Hatori84,BiskampWelter89,Zhou19}, which conserves `anastrophy', or the square of the magnetic vector potential (see Appendix \ref{sec:2d}), resulting in a $t^{-1}$ decay law independently of the reconnection regime. The evidence that has been presented for this picture in three dimensions relies on demonstrating that the mean-square magnetic vector potential (defined according to some particular, necessarily non-unique gauge choice) changes more slowly with time than does the magnetic energy~\cite{Bhat20}. However, this will be true for any decay satisfying $B^\alpha L\sim \const$ for any $\alpha>0$. Here, we propose a different theory of non-helical decay. 

\subsection{Qualitative theory of non-helical decay \label{sec:qualitative}}

The key point, already made in Section \ref{sec:intro}, is that a vanishing mean helicity density does not imply that individual magnetic structures are non-helical, because helicity is not a sign-definite quantity. Moreover, if not constrained by higher-order topological invariants, non-helical magnetic structures will relax to zero energy on the ideal timescale (as required by J.B.~Taylor relaxation \cite{Taylor74, Taylor86}). For example, consider a toroidal structure without a twist --- such a structure will shrink under the magnetic tension force (driving outflows along its axis) to zero magnetic energy. In contrast, for a toroidal structure with a net twist relative to the poloidal axis, such a relaxation is topologically impossible. Thus, we expect a collection of helical structures of different signs to be the natural state of non-helical MHD turbulence. Indeed, visualisations of our simulations do appear to support this intuition, see Fig.~\ref{fig:3d_slices}.

To motivate the more formal approach to follow, we first present an informal `cartoon' of the expected dynamics (Fig.~\ref{fig:cartoon}).  Consider a volume filled with helical magnetic structures, which we shall refer to as `blobs', as we wish to remain agnostic about their precise morphology. As in the helical case, we expect that topological constraints will hinder their relaxation on ideal timescales, and that instead the blobs will evolve via coalescences with other blobs on the prevailing reconnection timescale. 


For simplicity, suppose that all blobs have helicity $\sim H$, and for the moment, that they all have the same sign of helicity. When any two blobs merge, the resulting structure will have helicity $H' = 2H$, implying that the characteristic magnetic field and length scale, $B$ and $L$,  will satisfy $B'^2 L'^4 \sim 2 B^2 L^4$. If every blob in the system undergoes such a pairwise merger, then the total number of blobs, $N$, will decrease by a factor of $2$: $N'=N/2$. Assuming the blobs fill all space, their characteristic size must then increase as $L'=2^{1/3}L$. Together, these relations imply $B^2 L \sim \const$, which, of course, is precisely the condition obtained from the conservation of the volume-averaged helicity density, Eq.~\eqref{B2L}. 

In contrast, for a system with vanishing total helicity, there will be equal numbers of blobs with helicities $-H$ and $+H$. When two blobs with opposite helicities merge, the resulting structure will be non-helical and, assuming that no higher-order topological invariants constrain its subsequent evolution, it will relax to zero magnetic energy on the ideal timescale~\footnote{Even if the Lundquist number is very large and reconnection is plasmoid dominated and fast, there will still be a separation of timescales between the ideal timescale and the reconnection one, the latter being a factor of $10^2$ longer \cite{Uzdensky10}.}. In other words, when blobs of opposite helicity merge, they mutually `annihilate'. Otherwise, the signs of the helicities of the blobs should not modify the dynamics, so the system will have no preference between like-helicity and opposite-helicity mergers. This implies that, after one merger timescale, we will have $N'=N/4$, because of any four randomly chosen blobs, on average, two will annihilate, and two will merge to form a single blob. Again, assuming blobs fill all space, we have $L'=2^{2/3}L$, which, together with $B'^2 L'^4 \sim 2 B^2 L^4$, implies 
\begin{equation}
    B^4 L^5 \sim \const. \label{B45L}
\end{equation}This new scaling is the non-helical analogue of Eq.~\eqref{B2L}.

This picture, while conceptually simple, does rely on significant assumptions about the nature of the dynamics that are not obviously justifiable, e.g., that all structures have the same length scale and helicity, that the only dynamical process is mergers (rather than, say, fragmentation due to MHD instabilities) and that all mergers are pairwise processes. We now discuss how to formalise it. 

\subsection{Invariant-based theory of non-helical decay \label{sec:formal}}

\subsubsection{The Saffman helicity invariant}

We propose that the general evolution of a collection of localised structures of mixed magnetic helicity should conserve the integral
\begin{equation}
    I_H = \int \dd^3 \br \langle h(\bx) h(\bx+\br)\rangle, \label{saffman_helicity_invariant}
\end{equation}where, as before, $h=\bA \bcdot \bB$ is the helicity density, and angle brackets denote an ensemble average. The form of this integral is immediately reminiscent of the integrals that govern hydrodynamic decay: the already-introduced Loitsyansky integral, Eq.~\eqref{Loitsyansky}, and the Saffman integral~\cite{Saffman67},
\begin{equation}
    I_{\bP} = \int \dd^3 \br \langle \bu(\bx)\bcdot \bu(\bx+\br)\rangle. \label{saffman_integral}
\end{equation}The Saffman integral is finite for, and conserved by, hydrodynamic turbulence initialised with strong long-range spatial correlations, corresponding to a kinetic-energy spectrum $\propto k^2$ at the largest scales \cite{Saffman67} (we shall return to the connection with small-$k$ part of the spectrum below). Such turbulence is known as `Saffman turbulence' \cite{Davidson13}. When strong correlations are absent, $I_{\bP}=0$, and the Loitsyansky integral, Eq.~\eqref{Loitsyansky}, is conserved instead \cite{Saffman67, Ishida06, Davidson13}. This case is known as `Batchelor turbulence', after Batchelor and Proudman, who explored its properties in \cite{BatchelorProudman56}. Physically, the conservation of $I_{\bP}$ is related to the conservation of linear momentum, ${\bP \equiv \int \dd^3 \bx\, \bu}$, much as the conservation of the Loitsyansky integral is related to angular-momentum conservation, and, as we are about to show, $I_H$ is related to helicity conservation. Owing to this analogy, of which we shall make further profitable use in Section~\ref{sec:discussion}, we will refer to $I_H$ as the `Saffman helicity invariant'. We proceed by establishing the claims that, in non-helical turbulence, $I_H$ is gauge invariant, finite, and conserved. These arguments are in most respects analogous to those originally made by Saffman for $I_{\bP}$~\cite{Saffman67} (see \cite{Davidson13} for a review).
\newline
\newline
\textit{1. $I_H$ is gauge-invariant.}

Assuming that volume and ensemble averages are the same,
\begin{equation}
    I_H = \lim_{V \to \infty} \frac{1}{V} \left[ \int_V \dd^3 \bx \, h(\bx) \right]^2 = \lim_{V \to \infty} \frac{1}{V} 
    \langle H_V^2 \rangle, \label{shi_H^2/V}
\end{equation} so $I_H$ is the density of magnetic helicity squared. Gauge invariance is then guaranteed in the same manner as for the magnetic helicity, by arranging that the surface of the volume $V$ is always normal to the magnetic-field direction. This can always be achieved because of our assumption that the magnetic field forms localised structures, which are arbitrarily small compared to $V$ in the limit $V \to \infty$.
\newline
\newline
\textit{2. $I_H$ is finite.}

Returning to the definition of $I_H$, Eq.~\eqref{saffman_helicity_invariant}, and assuming that the system has no preference for accumulation of like- or opposite-helicity structures, $\langle h(\bx) h(\bx+\br)\rangle$ is zero if $\boldsymbol{r}$ extends beyond the characteristic size of a helical structure, $L$, and of size $\sim \langle h^2\rangle \sim B^4 L^2$ otherwise. Integrating over $\br$ then gives $I_H\sim B^4 L^5$. Formally, this requires the two-point magnetic-helicity-density correlation function to decay faster than $r^{-3}$ as $r\to\infty$. This is the condition for the magnetic structures to be `localised'. 

Another way to obtain this scaling is to consider the volume integral in Eq.~\eqref{shi_H^2/V} as a random walk in the net helicity contained within the volume $V$ --- the number of `steps' is $V/L^3$, so 
\begin{equation}
    H_V=\int_V \dd^3 \bx\, h(\bx) \sim \left(\frac{V}{L^3}\right)^{1/2}B^2 L^4. \label{random_walk}
\end{equation}Then there is cancellation of $V$ in Eq.~\eqref{shi_H^2/V}, and the scaling $I_H\sim B^4 L^5$ is recovered.
\newline
\newline
\textit{3. $I_H$ is conserved.}

According to Eq.~\eqref{random_walk}, the expectation value of the square helicity in a large volume of non-helical turbulence is $\langle H^2_V\rangle \propto V$. As $\eta \to 0^+$, the magnetic helicity is conserved during all processes that occur locally inside the volume $V$, so $\langle H^2_V\rangle \propto V$ can only be changed by processes occurring at the surface of $V$, $S=\partial V$. These are fluxes of helicity in or out of the volume, or else reconnection with magnetic structures not contained within the volume. However, both are random processes and hence the net rate of change of square helicity associated with them scales as $\sim S \sim V^{2/3}$. Ultimately, then, we find that
\begin{equation}
    \frac{1}{\langle H^2_V \rangle}\frac{\dd \langle H^2_V \rangle}{\dd t} \propto V^{-1/3},
\end{equation} with the result that in the limit $V \to \infty$, $\langle H_V^2\rangle$ is a conserved quantity. Therefore, so is $I_H$.

An alternative proof of the invariance of $I_H$, which follows directly from the MHD induction equation under the assumption of sufficiently rapid decay of long-range correlations, is detailed in Appendix~\ref{sec:app_invariance_proof}.

In the context of this work, the primary importance of the conservation of $I_H$ is that it implies precisely the same scaling as we found from our qualitative theory, Eq.~\eqref{B45L}. The more formal argument proposed here is much more general, however: while it requires that helical structures be localised (i.e., that local correlations decay sufficiently quickly with distance), and that there be no preference for accumulation of like-helicity or opposite-helicity structures, it does not require that all structures be of the same size or magnitude of helicity at any given instant, and neither does it require that the only relevant dynamics be pairwise mergers. The cartoon presented in Section \ref{sec:qualitative}, should, therefore, be considered as a particular example of dynamics that would conserve $I_H$, and that, therefore, must produce the correct scaling, but not as the only dynamics allowed in the system, or required to be prevalent in order for the scaling~\eqref{B45L} to hold.

\subsubsection{Permanence and impermanence of the large scales\label{sec:invariants+spectra}}

We now discuss some important consequences of the invariance of $I_H$ for the small-$k$ asymptotics of the spectra. In particular, we shall find that the phenomenon of `inverse transfer' of magnetic energy~\cite{Brandenburg15, ReppinBanerjee17}, whose explanation has so far been unclear, follows naturally from the conservation of $I_H$. We shall also find that the invariance of $I_H$ implies an invariant small-$k$ asymptotic of the helicity-variance spectrum, a fact that we shall utilise in Section~\ref{sec:nonhelical_results} to provide a numerical test of our theory.

As is well known, the Saffman and Loitsyansky integrals are, respectively, proportional to the coefficients of $k^2$ and $k^4$ in the small-$k$ asymptotic expansion of the kinetic-energy spectrum. To see why, note that if correlations between distant points decay sufficiently quickly, \textit{viz.}, if ${\langle \bu (\bx) \bcdot \bu (\bx + \br) \rangle < O(r^{-5})}$ as $r \to \infty$, then the kinetic-energy spectrum,
\begin{align}
    \mcE_K(k) & = \frac{k^2}{4\pi^2} \int \dd^3 \br\, \langle \bu (\bx) \bcdot \bu (\bx + \br) \rangle e^{-i\bk \bcdot \br},
    \label{E(k)exact}
\end{align}may be Taylor-expanded in $kL\ll 1$, which yields (under the assumptions of statistical isotropy and homogeneity),
\begin{equation}
    \mcE(k\to 0) = \frac{I_{\bP} k^2}{4 \pi ^2} + \frac{I_{\bL} k^4}{24 \pi^2} + O(k^5). \label{Eexpansion}
\end{equation}
Thus, Saffman turbulence, with $I_{\bP} \neq 0$, has ${\mcE (k\to 0)\propto k^2}$ \cite{Saffman67}, while Batchelor turbulence, with $I_{\bP} = 0$, has ${\mcE (k\to 0)\propto k^4}$ instead \cite{BatchelorProudman56}.
Owing to the invariance of $I_{\bP}$ and $I_{\bL}$, Eq.~\eqref{Eexpansion} leads to a phenomenon known as the `permanence of the large-scale eddies' --- as hydrodynamic turbulence decays, the small-$k$ part of its energy spectrum remains unchanged. In particular, this means that non-helical hydrodynamic turbulence supports no inverse energy transfer. This is unlike non-helical MHD, which, if initialised with ${\mcE (k\to 0)\propto k^4}$ (or steeper, like in our simulations --- see Appendix~\ref{sec:app_plaws}), has been found in simulations to increase its energy content at large scales~\cite{Brandenburg15, ReppinBanerjee17}. This may be interpreted as a consequence of the non-invariance of the magnetic equivalent of the Loitsyansky integral,
\begin{equation}
    I_{\bL_{M}}\equiv-\int\dd^3 \br \,r^2 \langle \bB(\bx)\bcdot\bB(\bx+\br) \rangle,
\end{equation}which is related to the magnetic energy spectrum via the expansion analogous to Eq.~\eqref{Eexpansion},
\begin{equation}
    \mcE_{M}(k\to 0) = \frac{I_{\bB} k^2}{4 \pi ^2} + \frac{I_{\bL_{M}} k^4}{24 \pi^2} + O(k^5), \label{EMexpansion}
\end{equation}where $I_{\bB}$ will be defined and discussed in Section~\ref{sec:Saffman_fails}, but for now can be assumed to be zero.

Of course, there was no reason to suspect that $I_{\bL_{M}}$ should have been a dynamical invariant, as angular momentum does not have a magnetic equivalent --- ${\bL_{M}\equiv\br \times \bB}$ is not a conserved quantity in MHD. In fact, the growth of $I_{\bL_{M}}$, and hence the inverse transfer, can be recovered immediately from the conservation of the Saffman helicity invariant, as
\begin{equation}
    I_{\bL_{M}}\sim B^2 L^5 \sim \frac{I_H}{B^2},\label{inverse_transfer}
\end{equation}so if $I_H$ is conserved while $B^2$ decays, $I_{\bL_{M}}$ must grow.

Owing to the presence of the inverse transfer, it would appear that there is no `permanence of the large scales' principle for non-helical MHD. However, a modified version of this principle may be obtained by noting that $I_H$ is proportional to the coefficient of $k^2$ in the small-$k$ expansion of the helicity-variance spectrum,
\begin{align}
    \Theta(k) & = \frac{k^2}{2\pi^2} \int \dd^3 \br\, \langle h (\bx)\, h (\bx + \br) \rangle e^{-i\bk \bcdot \br}.
    \label{Theta(k)exact}
\end{align}Namely, at small $k$,
\begin{equation}
    \Theta(k\to 0) = \frac{I_{H} k^2}{2 \pi ^2} + O(k^3), \label{Thetaexpansion}
\end{equation}provided that $\langle h (\bx)\, h (\bx + \br) \rangle<O(r^{-3})$ as $r\to\infty$. Thus, non-helical MHD does have a kind of `permanence of the large scales' phenomenon, though it manifests itself in the helicity-variance spectrum, not in the energy spectrum. Detecting this phenomenon numerically will be a useful test of our theory and a way of confirming the conservation of $I_H$ (see Section \ref{sec:nonhelical_results}).

\subsection{Decay laws \label{nonhelical_laws}}

Let us now compute the decay laws associated with the conservation of $I_H$. Eq.~\eqref{reconnecting_plaw} with $\alpha = 4/5$, as demanded by Eq.~\eqref{B45L}, implies a magnetic-energy decay
\begin{equation}
    E_M \propto t^{-20/17} \simeq t^{-1.18},
\end{equation}if the dynamics occur on the Sweet-Parker timescale (with Laplacian viscosity), or Eq.~\eqref{ideal_law} gives
\begin{equation}
    E_M \propto t^{-10/9} \simeq t^{-1.11},
\end{equation}if reconnection is fast, i.e., either stochastic or plasmoid-dominated. These exponents are both close to $-1$, so are consistent with previous numerical results that have reported a power law decay $\propto t^{-1}$ \cite{Zrake14, Brandenburg15, Brandenburg17,ReppinBanerjee17, Bhat20}.

\begin{figure}
\includegraphics[width=\columnwidth]{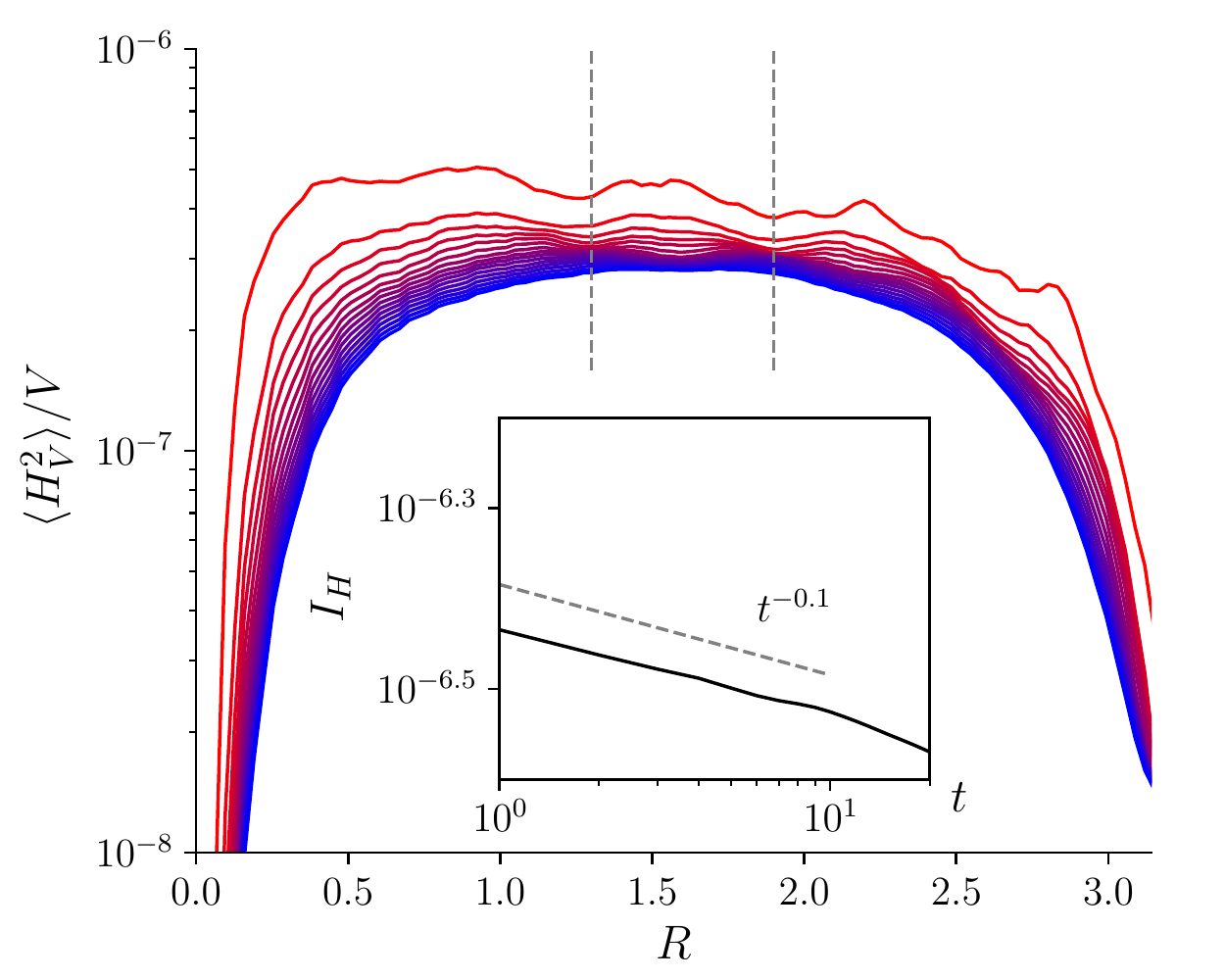}
\caption{\label{fig:saffman_helicity_invariant}Confirmation of the expected scaling of $\langle H_V^2 \rangle/V$ with volume, $V= (2R)^3$, for a simulation with $n=6$, $\eta_6 = 1.42 \times 10^{-12}$. Flat parts of the curves correspond to the volume-independent limit, as expected by Eq.~\eqref{random_walk}. Curves are plotted with a constant interval of $1.0$ between $t=0.0$ (red) and $t=15.0$ (blue), where time is in code units, as explained in the caption to Fig.~\ref{fig:helical_spectrum}.
The inset shows the evolution of $I_H$, computed as the mean value of $\langle H_V^2 \rangle/V$ between $R=1.3$ and $R=1.9$.}
\end{figure}

\begin{figure}
\includegraphics[width=\columnwidth]{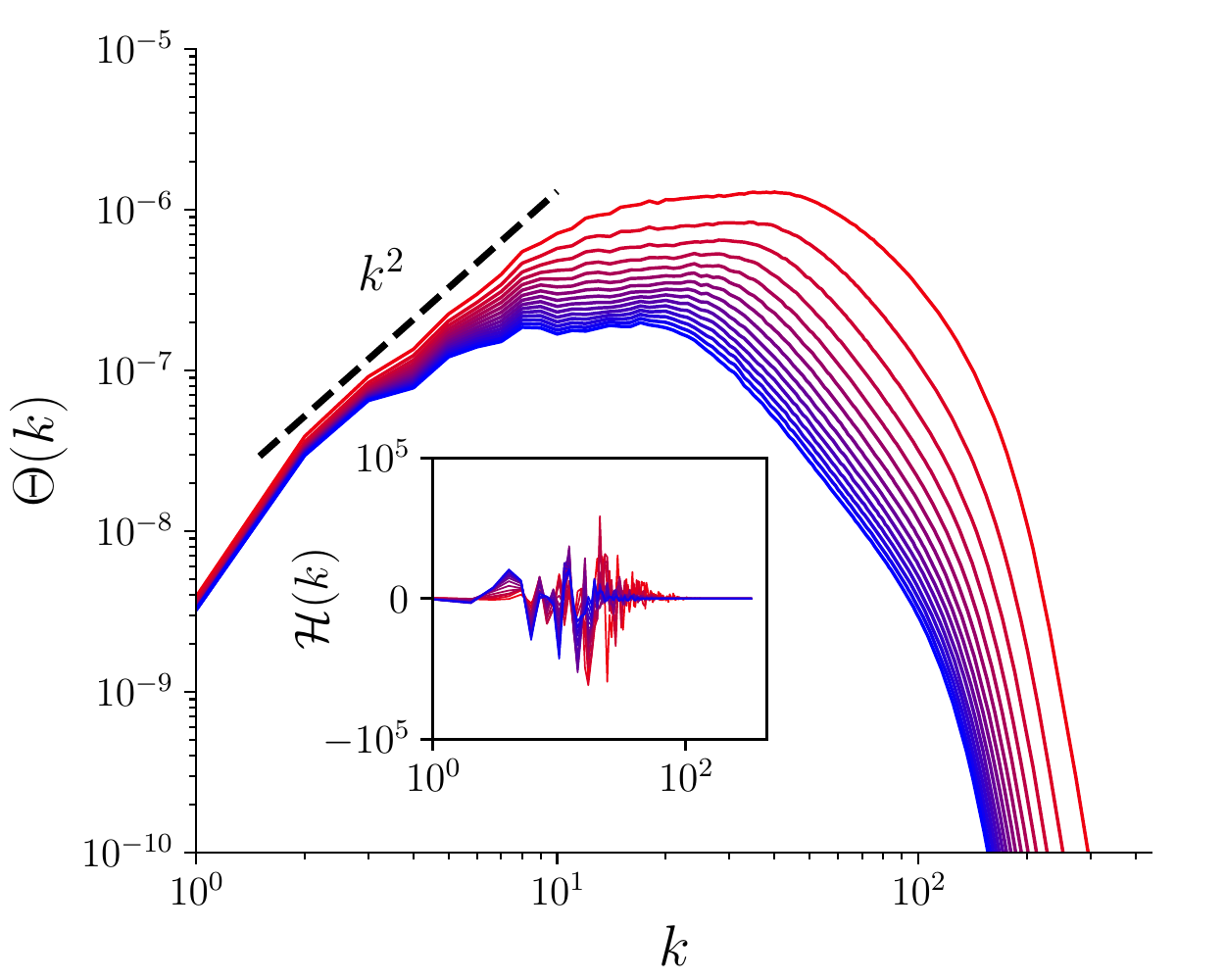}
\caption{\label{fig:helicity_spectrum}The evolution of the helicity-variance spectrum, Eq.~\eqref{Theta(k)exact},  plotted for the same non-helical simulation and at the same times as Fig.~\ref{fig:saffman_helicity_invariant}. An invariant small-$k$ asymptotic $\Theta(k)\propto k^2$ indicates finiteness and conservation of $I_H$, as explained in Section~\ref{sec:invariants+spectra}. Inset: The helicity spectrum, plotted at the same times, for reference (with linear vertical axis). While small fluctuations are present, there is no strong net helicity of either sign at any scale.}
\end{figure}

For the kinetic energy, we again expect Eq.~\eqref{EK_sim_dL_EM} to hold, provided the kinetic energy is dominated by reconnection outflows. Under the Sweet-Parker scalings and $B^4 L^5 \sim \const$, this becomes
\begin{equation}
    E_K\sim E_M^{(14n-9)/10n}.\label{nonhelical_KE}
\end{equation}For $n=2$ and $n=4$, this gives $E_K\sim E_M^{19/20}$ and ${E_K\sim E_M^{47/40}}$, respectively. The closeness of these exponents to~$1$ indicates that the current-sheet aspect ratio changes more slowly in the non-helical case than the helical case, and explains why no significant difference in the kinetic- and magnetic-energy decay laws has been reported in the previous numerical studies cited above.


\subsection{Numerical results\label{sec:nonhelical_results}}

To test the theory proposed in Sections~\ref{sec:formal} and~\ref{nonhelical_laws}, we now present results from simulations of decaying non-helical turbulence, analogous to those presented for helical fields in Section~\ref{sec:helical_results}.

We first address the question of the scaling and conservation of $I_H$ in our simulations. In a periodic box, rather than infinite space, one should interpret the limit $V\to\infty$ in Eq.~\eqref{shi_H^2/V} as requiring $V$ to be large compared to the energy-containing scales, but small compared to the box size, where the assumption of isotropy fails, as does the random-walk scaling of magnetic helicity, Eq.~\eqref{random_walk}, if the total helicity in the box is constrained to be exactly zero by the initial condition.  Fig.~\ref{fig:saffman_helicity_invariant} shows plots of $\langle H_V^2 \rangle/V$ vs. $R$, where $V$ is a cube of width $2R$, and we take an ensemble average over many different cube positions in the simulation box (employing the Coulomb gauge, $\nabla \bcdot \bA=0$, for numerical convenience). As $R\to 0$, $\langle H_V^2 \rangle/V\propto V\to 0$, because $\langle H_V^2 \rangle$ is dominated by individual structures. Similarly, $\langle H_V^2 \rangle/V$ vanishes as $R\to \pi$, because then $V$ is the entire periodic simulation domain, which has zero magnetic helicity by construction. However, for intermediate values of $R$, $\langle H_V^2 \rangle/V$ turns out to be independent of $V$, confirming the random-walk scaling, Eq.~\eqref{random_walk}.

\begin{figure}
\includegraphics[width=\columnwidth]{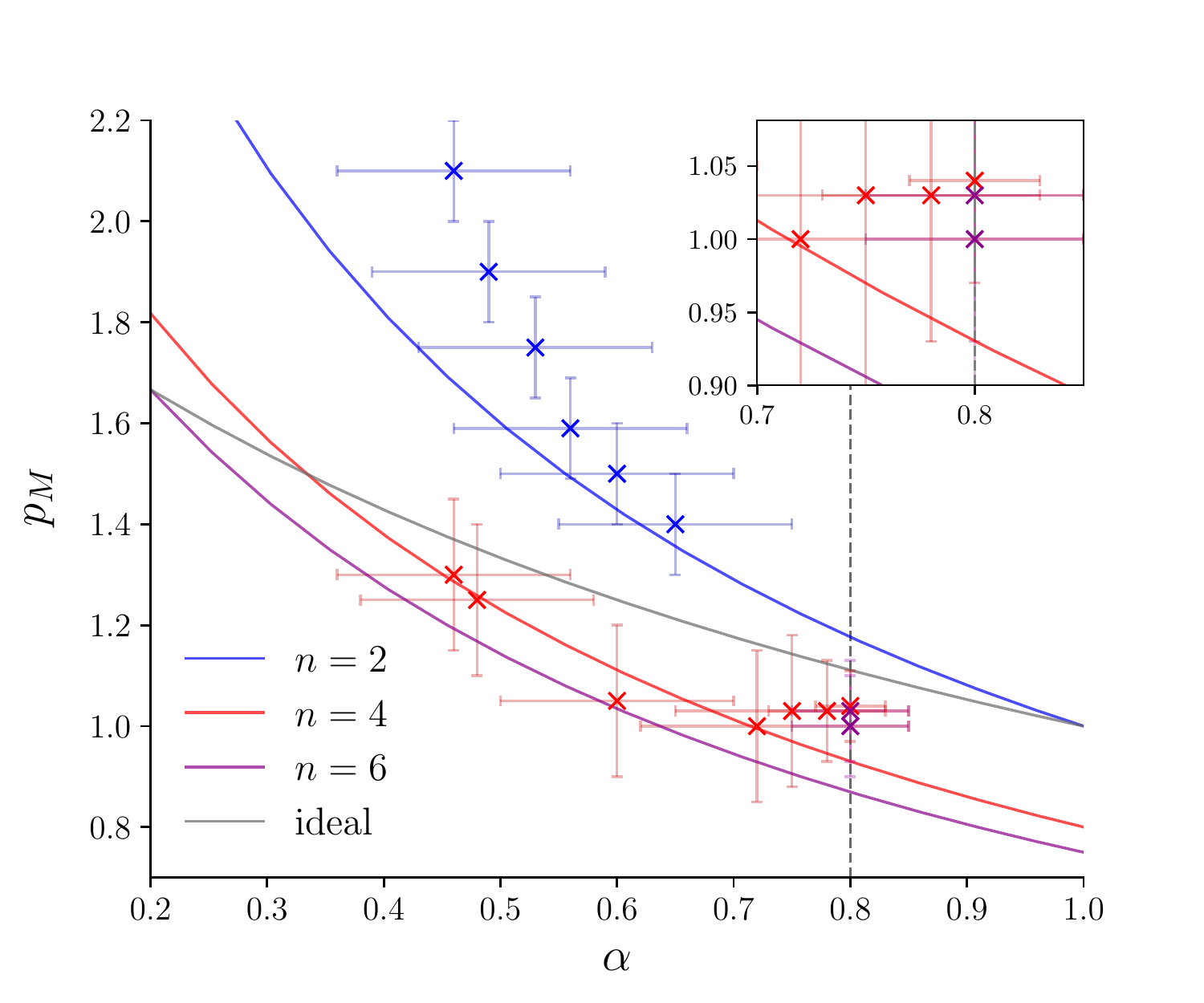}
\caption{\label{fig:p_alpha_nonhelical}Relation between the empirically obtained magnetic-energy-decay power-law exponent, $p_M$, and the value of $\alpha$ for which $B^\alpha L \sim \const$ for the non-helical simulations. As in Fig.~\ref{fig:p_alpha_helical}, solid curves show the expected relationship, Eq.~\eqref{reconnecting_plaw}, for decays occurring on the Sweet-Parker timescale, with $n=2$, $n=4$ and $n=6$ shown in blue, red, and magenta respectively. The grey solid curve depicts the scaling given by Eq.~\eqref{ideal_law}. Simulation results are in excellent agreement with the coloured curves, with better agreement as $\alpha$ increases towards the limiting value of $4/5$. As in the helical case, simulations at the largest Lundquist numbers appear to be on the brink of the transition to the fast-reconnection regime. The full set of decay curves from which this plot was derived can be found in Appendix~\ref{sec:app_plaws}. }
\end{figure}

The value of $I_H$ is $\langle H_V^2 \rangle/V$ in this flat region. Computing it as the average of $\langle H_V^2 \rangle/V$ between the two dashed lines in Fig.~\ref{fig:saffman_helicity_invariant} as a function of time, we find that $I_H$ decays approximately as $t^{-0.1}$ (see the inset to Fig.~\ref{fig:saffman_helicity_invariant}). Considering the strong scaling of $I_H$ with $B$ and $L$ --- $I_H\sim B^4 L^5$ --- this decay is very slow, i.e., Eq.~\eqref{B45L} holds well.
For comparison, if anastrophy were conserved, as in two dimensions, then, according to the robust Sweet-Parker scaling for $n=4$, viz., $BL\sim \const$, $B^2 \sim t^{-3/4}$ (see Appendix~\ref{sec:2d}), $I_H$ would  \textit{grow}: $B^4 L^5 \sim t^{3/8}$ (and, indeed, even faster growth should be expected under the often-assumed $B^2 \sim t^{-1}$, $L\sim t^{1/2}$, since then $B^4 L^5 \sim t^{1/2}$).
Thus, the distinction between our theory of non-helical decay and the conjecture of quasi-two-dimensional dynamics \cite{Bhat20, Brandenburg15, Brandenburg17} is measurable in numerical simulations, and there is strong evidence in support of the former over the latter.

The conservation of $I_H$ may also be demonstrated from the invariance of the small-$k$ part of the helicity-variance spectrum, $\Theta(k)$, as explained in Section~\ref{sec:invariants+spectra}. The evolution of this spectrum is shown in Fig.~\ref{fig:helicity_spectrum}. The small-$k$ part exhibits a $k^2$ power law, indicating that the expansion Eq.~\eqref{Thetaexpansion} is valid, and that $I_H$ is finite. As the turbulence decays, the small-$k$ part is preserved to good approximation, though there is a small amount of decay, owing to the finite scale separation between the box scale (close to which, $\Theta(k)\propto k^2$ should fail) and the energy-containing scales. Nonetheless, the behaviour is once again markedly different from what should be expected under the often-assumed $B^2 \sim t^{-1}$, $L\sim t^{1/2}$ laws. In that case, an inverse transfer of helicity variance should be expected, with the small-$k$ part of $\Theta(k)$ growing like $B^4 L^5\sim t^{1/2}$.

\begin{figure}
\includegraphics[width=\columnwidth]{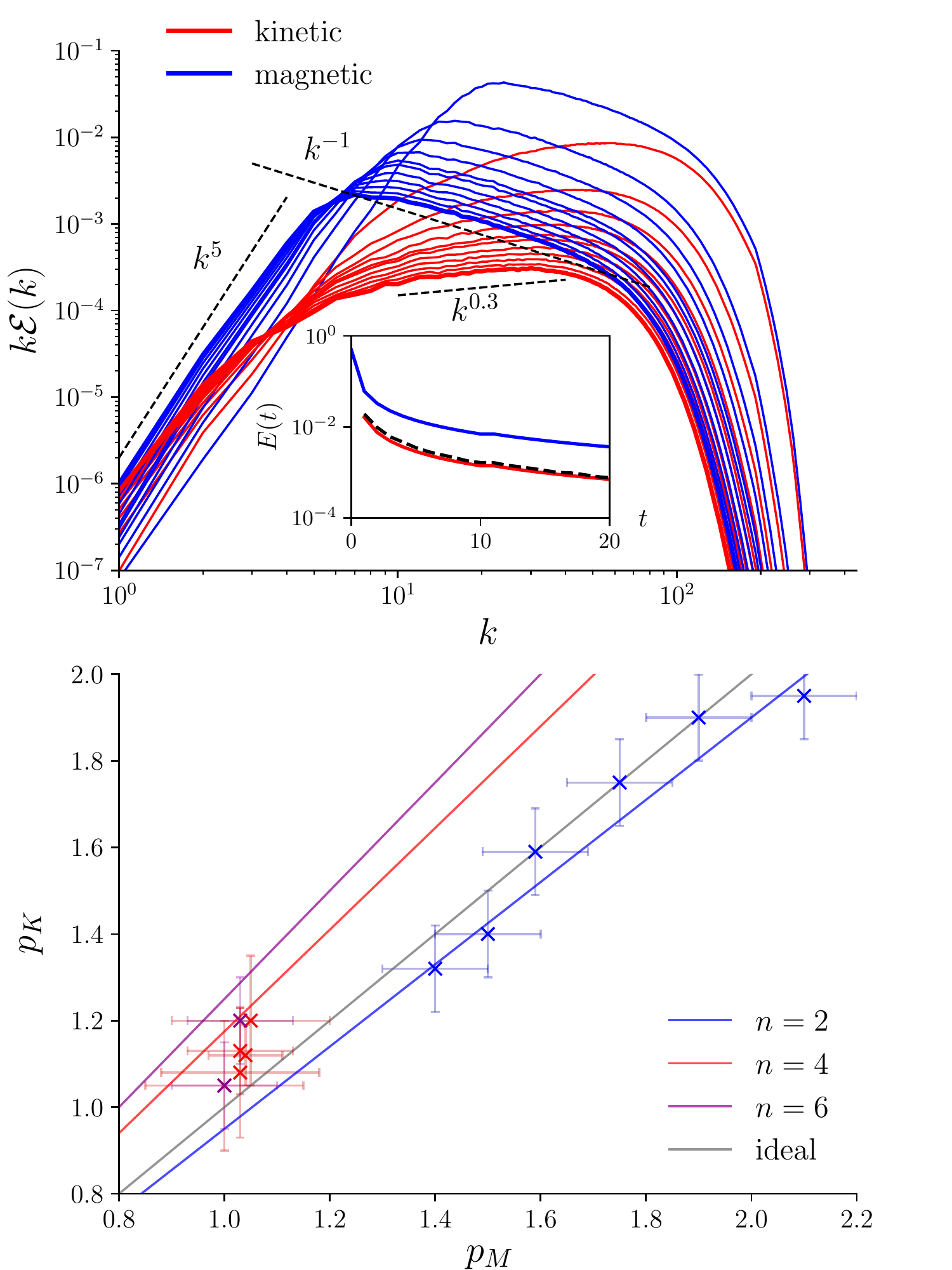}
\caption{\label{fig:nonhelical_spectrum} Top panel: Evolution of $k\mcE(k)$, where $\mcE(k)$ is the spectral energy density, for the magnetic (blue) and kinetic (red) energies. These plots are obtained from the non-helical simulation with $n=4$, $\eta_4 = 2\times 10^{-8}$. Each plot of $k\mcE(k)$ is separated by a time interval of $2.0$ between $t=1.0$ to $t=21.0$ ($t=21.0$ in bold), where time is in code units, as explained in the caption to Fig.~\ref{fig:helical_spectrum}. As in the helical case (Fig.~\ref{fig:helical_spectrum}), the peak of the magnetic energy is at much larger scales (smaller $k$) than the peak of the kinetic-energy spectrum, consistent with the expectation that the kinetic energy should be contained within the Sweet-Parker sheets. Inset: decay of the total magnetic and kinetic energies. The total kinetic energy curve is much below the magnetic energy curve, and coincides with $(\delta/L)E_M$ (black), in agreement with Eq.~\eqref{EK_sim_dL_EM}. Here, we compute $\delta/L$ as the ratio of the wavenumbers at which $k\mcE_M(k)$ and $k\mcE_K(k)$ peak.
Bottom panel: The kinetic-energy decay exponent, $p_K$, against the magnetic-energy decay exponent, $p_M$, as measured in simulations with $S_{n,\,0}^{1/n}>9.0$. Results are in reasonable agreement with the theoretical prediction, Eq.~\eqref{nonhelical_KE} (coloured lines), though as noted in the main text, this prediction is very similar to $E_K\propto E_M$ (grey).}
\end{figure}

\begin{figure*}
\includegraphics[width=\textwidth]{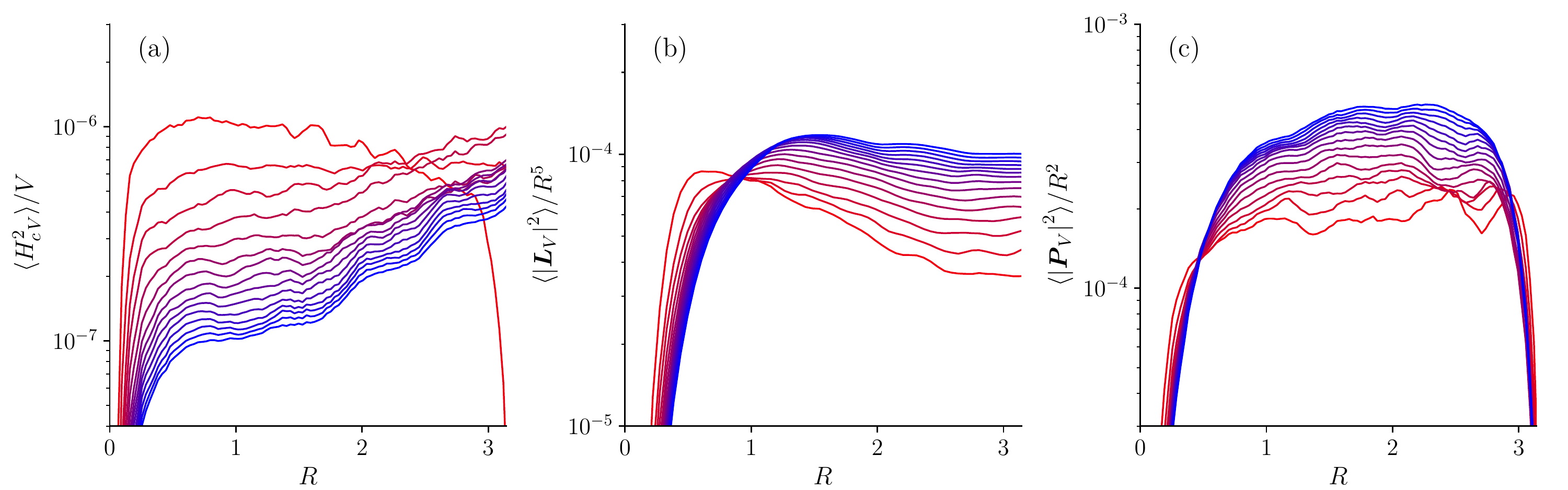}
\caption{\label{fig:other_saffman}Evolution of Saffman-type invariants (other than the helicity invariant) in a non-helical simulation with $n=6, \,\eta_6=1.42\times 10^{-12}$. Curves are plotted with a constant interval of $1.0$ between $t=1.0$ (red) and $t=15.0$ (blue), where time is in code units, as explained in the caption to Fig.~\ref{fig:helical_spectrum}. (a) $\langle H_{c\,V}^2\rangle$, where $H_{c\,V}$ is the cross-helicity contained in a cube with~$V=(2R)^3$. While the expected random-walk scaling $\langle H_{c\,V}^2\rangle\sim V$ is obeyed, $\langle H_{c\,V}^2\rangle$ is not conserved. (b) $\langle {|\bL_V|}^2\rangle$, where $\bL_V$ is the angular momentum contained in a \textit{sphere} of radius $R$, about the centre of the sphere. We find $\langle {|\bL_V|}^2\rangle\sim R^5$, at late times, though the scaling is closer to $R^4$ initially. (c) $\langle {|\bP_V|}^2\rangle$, where $\bP_V$ is the linear momentum contained in a cube with $V=(2R)^3$. We find $\langle {|\bP_V|}^2\rangle\sim R^2$, though the scaling appears to become somewhat stronger with time.}
\end{figure*}

Turning to the measured decay laws, we note that, as in the helical case, the asymptotic scaling $B^4 L^5\sim \const$ will not necessarily be satisfied for a decay at finite $\eta_n$. Nevertheless, we can still measure a value of $\alpha$ for which $B^{\alpha}L\sim \const$ is satisfied, and compare the measured value of the magnetic-energy decay exponent, $p_M$, with the one expected for a decay on Sweet-Parker or ideal timescales. The results of this comparison are shown in Fig.~\ref{fig:p_alpha_nonhelical}. While agreement with the Sweet-Parker curves (coloured) for $\alpha < 4/5$ is not quite as good as in the helical case, we still observe (\textit{i}) a clear preference for the Sweet-Parker prediction over the prediction of a decay on ideal timescales (grey), and (\textit{ii}) convergence to $B^{4/5}L\sim \const$ for the hyper-dissipative simulations.

Finally, we describe the decay of kinetic energy. Fig.~\ref{fig:nonhelical_spectrum} again shows that, like in the helical case, the kinetic energy is peaked at smaller scales than the magnetic energy is, consistent with the expectation of Alfv\'{e}nic outflows in current sheets. We also find that Eq.~\eqref{EK_sim_dL_EM} is very well satisfied, and a reasonable agreement with Eq.~\eqref{nonhelical_KE}, although the magnetic- and kinetic-energy decay exponents are very close to each other.

\subsection{The behaviour of other invariants \label{sec:otherinvariants}}

To conclude the discussion of numerical results, we now address the evolution of the other, better known invariants during our simulations, namely, the cross-helicity, as well as the Loitsyansky and Saffman integrals.

First, we consider the cross-helicity, 
\begin{equation}
    H_c = \int \dd^3 \br\, \bu \bcdot \bB,
\end{equation}which is an ideal invariant of the (incompressible) MHD equations, though we find that it is not conserved in our simulations any better than energy is. In Fig.~\ref{fig:other_saffman}(a), we plot the average value of the squared total cross-helicity contained in a cube of volume $V=(2R)^3$, in a manner analogous to Fig.~\ref{fig:saffman_helicity_invariant} for the magnetic helicity. We find that $\langle H_{c\,V}^2 \rangle \sim V$ for $R \lesssim \pi/2$, as is expected from the random-walk argument. For $R> \pi/2$, $\langle H_{c\,V}^2 \rangle / V$ increases with $R$ but not as fast as $R^3$, which indicates that there is no net cross-helicity in the box. Even so, one may consider the Saffman-type `invariant', $I_{H_c}$, that is associated with cross-helicity (see Sections~\ref{sec:meanfield} and \ref{sec:UsimB}). It is given by the value of $\langle H_{c\,V}^2 \rangle / V$ in the flat region of Fig.~\ref{fig:other_saffman}(a), and, as we see, decays. The reason for this behaviour is that the cross-helicity's decay rate has the same scaling with the dissipation coefficients $\eta_n$ (resistivity) and $\nu_n$ (viscosity) as the decay rate of the magnetic energy does (and as is inevitable, because cross-helicity and energy have the same physical dimensions).

Likewise, the Loitsyansky integral, which encodes the statistics of angular-momentum fluctuations~\cite{LandauLifshitzFluids, Davidson09}, is not conserved in our simulations, as is clear from the small-$k$ part of Fig.~\ref{fig:nonhelical_spectrum}. In Fig.~\ref{fig:other_saffman}(b), we plot $\langle|\bL_V|^2\rangle$ against $R$, where $\bL_V$ is the total angular momentum contained in a spherical control volume $V$ of radius $R$, calculated about its centre. Presumably, this is due to injection of angular-momentum fluctuations by the reconnection outflows. Intriguingly, we observe a shift in the scaling properties of $\langle|\bL_V|^2\rangle$ vs. $R$ --- at early times, $\langle|\bL_V|^2\rangle \sim R^4$, which is the expected scaling for `Batchelor turbulence', i.e., when correlations between distant points are weak \cite{BatchelorProudman56, Davidson13}; subsequently, the system appears to evolve towards a state with $\langle|\bL_V|^2\rangle \sim R^5$. The latter is the `Saffman-turbulence' scaling, and suggests strong long-range correlations in the velocity field \cite{Davidson13, Saffman67}. A corresponding shift in the analogous quantity for linear momentum, $\langle|\bP_V|^2\rangle$, vs. $R$ is suggested by Fig.~\ref{fig:other_saffman}(c), which shows a stronger scaling than $R^2$ (the Batchelor scaling) for $R<\pi/2$ at later times, closer to the Saffman-turbulence scaling of $R^3$. This is also consistent with Figs. \ref{fig:helical_spectrum} and \ref{fig:nonhelical_spectrum}, which appear to show a decreasing slope in the large-scale kinetic-energy spectrum over time, perhaps towards $\mcE_K \propto k^2$, the hallmark of Saffman turbulence \cite{Saffman67, Davidson13}. This spectral behaviour has been noted in other studies \cite{Brandenburg15, Brandenburg17, Brandenburg19}, though it was considered an effect of compressibility, owing to the fact that the incompressible simulations of \cite{BereraLinkmann14} appeared not to see it. However, this may just have been because there was not enough time for the $k^2$ velocity spectrum to establish itself before the outer scale of the turbulence reached the box size in that study. While not present in decaying hydrodynamic turbulence \cite{Davidson13}, we suggest the effect might be related to the `thermalisation' phenomenon that is observed in forced, hydrodynamic turbulence \cite{Dallas15, Cameron17, AlexakisBiferale18, AlexakisBrachet19}. We shall address this topic specifically in a future publication~\cite{HoskingSchekochihin21k2}.

Finally, for the reader concerned that the standard scalings for $\langle|\bL_V|^2\rangle$ and $\langle|\bP_V|^2\rangle$ referred to here are not the same as the $\propto R^3$ random-walk scalings assumed in Section \ref{sec:formal}, we show how these scalings may be obtained from the random-walk approach in Appendix \ref{sec:LoitSaffDiscuss}. A more formal derivation of them may be found in \cite{Davidson13}.

\section{Discussion \label{sec:discussion}}

\subsection{Case of small, but non-zero, helicity\label{sec:fractional}}

In Section~\ref{sec:non-helical}, we proposed a way to impose the constraint of magnetic-helicity conservation on the decay laws of non-helical MHD turbulence, via the conservation of $I_H$.
Of course, no real field configuration will have precisely zero helicity, and therefore it is important to consider the evolution of a field configuration with small, but non-zero, magnetic helicity. In such a case, the system will undergo a transient (though, perhaps, long) period of evolution according to the non-helical decay law $B^4 L^5\sim \const$ [Eq.~\eqref{B45L}], before ultimately entering the fully helical regime, with a corresponding change in the decay law to $B^2 L\sim \const$ [Eq.~\eqref{B2L}].

This conclusion is an immediate consequence of the non-helical decay laws, as follows. Suppose that the system starts with some small helicity fraction $\sigma_0\ll 1$, defined so that the total helicity is $H=\sigma_0 B_0^2 L_0 V$. At a later time, since helicity is conserved,
\begin{equation}
    \sigma B^2 L = \sigma_0 B_0^2 L_0. \label{f_evolution}
\end{equation}Because $\sigma_0 \ll 1$, the system is not controlled by its total helicity, at least initially. It therefore evolves according to $B^4 L^5 \sim \const$. Using this in Eq.~\eqref{f_evolution}, we find 
\begin{equation}
    \sigma \sim  \sigma_0 \left(\frac{B}{B_0}\right)^{-6/5} \sim \sigma_0 \left(\frac{L}{L_0}\right)^{3/2}.
\end{equation}Thus, the helicity fraction $\sigma$ will grow with time. This can continue until $\sigma\sim1$, at which point the system enters the helical regime. This occurs when 
\begin{equation}
    B\sim B_0 \sigma_0^{5/6}, \quad L\sim L_0 \sigma_0^{-2/3}, \label{changeover_I_H_to_H}
\end{equation}provided that the energy-containing scale $L$ has not yet reached the system size.

This result is intuitive from the cartoon picture presented in Section~\ref{sec:non-helical}: when blobs with one sign of helicity are more populous, the ultimate consequence of random mergers is for the less populous type to be used up (although this can take a long time, which scales with an appropriate negative power of $\sigma_0$, if the initial population imbalance is small). 

\begin{figure}
\includegraphics[width=\columnwidth]{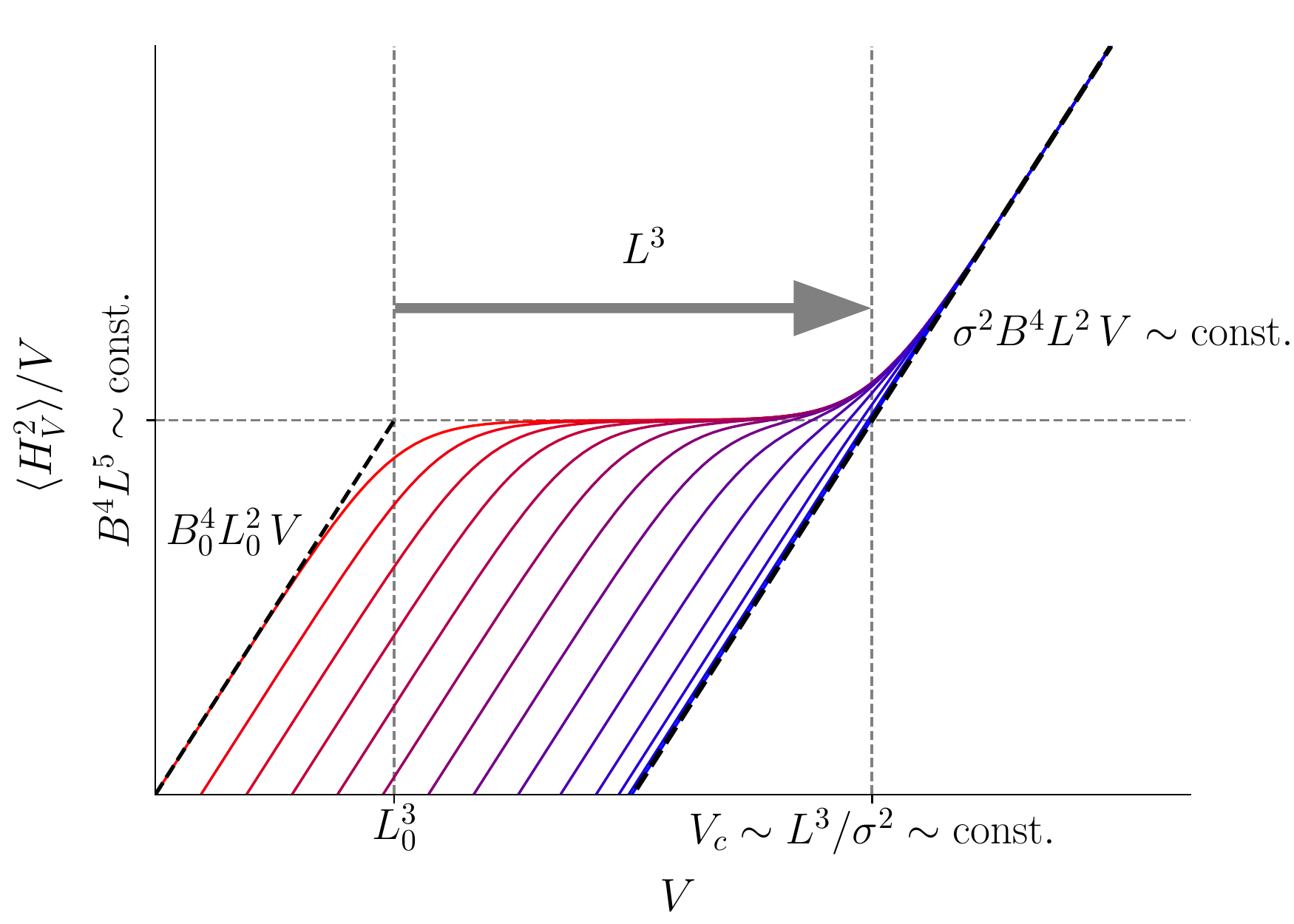}
\caption{\label{fig:SHI_to_helical_schematic} Schematic of $\langle H_V^2 \rangle/V$ as function of $V$, as a system with small initial fractional helicity, $\sigma_0$, transitions to the fully helical regime, as explained in the main text. The progression of time is shown by red~$\to$~blue, with plots at logarithmically spaced time intervals. Note that both axes are plotted on logarithmic scales.}
\end{figure}

The same conclusion can be reached from the consideration of the Saffman helicity invariant, Eq.~\eqref{shi_H^2/V}, though some care should be taken, as $I_H$ is formally infinite in the presence of any net helicity. However, if the helicity fraction $\sigma$ is small, then we can interpret the limit $V \to \infty$ in Eq.~\eqref{shi_H^2/V} as requiring $L^3 \ll V \ll V_c$, where $V_c$ is the critical volume at which $H_V$ ceases to be dominated by the net helicity fluctuation owing to its collection of helical structures of random signs, and instead is dominated by the helicity imbalance (see Fig.~\ref{fig:SHI_to_helical_schematic}).
This condition implies 
\begin{equation}
    \sigma \langle h^2 \rangle^{1/2}\, V_c \sim \langle h^2 \rangle^{1/2} L^3 \left(\frac{V_c}{L^3}\right)^{1/2} \implies V_c\sim \frac{L^3}{\sigma^2}.
\end{equation}For any $V$ such that $L^3 \ll V \ll V_c$, the arguments presented in Section~\ref{sec:formal} in favour of the conservation of $H_V$ remain valid and hence $I_H$ still provides the dominant constraint on the decay of magnetic structures. Choosing instead $V>V_c$, $\langle H_V^2 \rangle \sim \sigma^2 V B^4 L^2 \sim \const$, so we recover the evolution equation for $\sigma$, Eq.~\eqref{f_evolution}. However, when ultimately $\sigma \sim 1$, there is no longer any possibility of satisfying $L^3 \ll V \ll V_c$, because $V_c \sim L^3$. At this point, $\langle H_V^2 \rangle \sim V B^4 L^2$ for any chosen volume, and we are back to the fully helical scaling, Eq.~\eqref{B2L}.

These arguments suggest that the non-helical decay is ultimately transient for any real system, and there will always eventually be a transition to the helical regime, provided that the growing, energy-containing scale does not reach the system size before this happens. The same conclusion was reached by \cite{BanerjeeJedamzik04}, 
although their argument was based on different non-helical decay laws than those that we have proposed. A numerical simulation demonstrating the expected transition between the two regimes would be highly desirable, but imposes considerable numerical cost, so is left for future work.



\subsection{General decay principles \label{sec:general_principles}}

Let us now discuss how the principles that guided us in the above might be applied to other types of decaying turbulence. In any type of turbulence with an ideal invariant, $F=\int \dd^3 \bx \,f(\bx)$, that is better conserved than the energy, the conservation of that invariant implies 
\begin{align}
    \langle f \rangle = & \lim_{V \to \infty} \frac{1}{V}\int_V \dd^3 \bx\, f (\bx) \sim \const.\label{discussion_invariant}
\end{align}This is, in general form, the `selective decay' principle of \cite{Taylor74, Montgomery78, MatthaeusMontgomery80, Riyopoulos82, MatthaeusMontgomery84, Taylor86, Ting86}. 

If nonlinear structures (eddies, blobs) are necessarily in possession of $F$, so that the characteristic size of $f$ can be related to the sizes and correlation scales of the dynamical fields, then Eq.~\eqref{discussion_invariant} imposes a constraint on these fields that must be satisfied as the turbulence decays. In particular, if the sign of $f(\bx)$ within each structure is the same, either because it is sign-definite, or because the initial condition stipulates a predominance of structures of one particular sign, then $\langle f \rangle$ can be related to the characteristic sizes of dynamical fields and their correlation scales:
\begin{equation}
    \langle f \rangle \sim \psi^a L_\psi^b\,  \ldots, \label{scaling}
\end{equation}where $\psi$ is a representative dynamical field, $L_\psi$ is its correlation scale, $a$ and $b$ are exponents determined by the functional form of $f$, and `$\ldots$' denotes the possibility of other fields and scales. Together, Eqs.~\eqref{discussion_invariant} and~\eqref{scaling} imply a constraint on the dynamical fields that must be satisfied during the decay,
\begin{equation}
    \psi^a L_\psi^b\,  \ldots \sim \const.\label{constraint}
\end{equation}

Alternatively, for sign-indefinite $f$, there may be no strong predominance of structures associated with either sign of the invariant. In this case, 
\begin{equation}
    \langle f \rangle \sim \sigma \langle f^2\rangle^{1/2} \sim \sigma \psi^a L_\psi^b\,  \ldots, \label{fractional_scaling}
\end{equation}where $\sigma \ll 1$ is the fractional imbalance in $F$, as in Section~\ref{sec:fractional}. While Eq.~\eqref{discussion_invariant} still holds, its utility is reduced as $\sigma$ is generally a function of time --- the relationship 
\begin{equation}
    \sigma \langle f^2 \rangle^{1/2} \sim \sigma\psi^a L_{\psi}^b\dots\sim \const, \label{fractional_constraint}
\end{equation}implied by Eqs.~\eqref{discussion_invariant} and~\eqref{fractional_scaling}, should be considered as an evolution equation for $\sigma$, rather than a constraint on $\psi$ and $L_{\psi}$.

However, all is not lost. We propose that when $\sigma \ll 1$, conservation of \textit{local fluctuations} in $F$ imposes a constraint on the decay, through the associated Saffman-type integral
\begin{equation}
    I_F = \int \dd^3 \br\, \langle f(\bx) f(\bx+\br) \rangle \sim \langle f^2 \rangle L_f^3 \sim \const, \label{discussion_invariant_saffman}
\end{equation}where $L_f$ is the correlation scale of $f$. Again, if nonlinear structures are in possession of $F$, then $\langle f^2\rangle L_f^3$ can be related to the characteristic sizes and correlation lengths of dynamical fields,
\begin{equation}
    \langle f^2\rangle L_f^3 \sim \psi^c L_\psi^d\,  \ldots, \label{balanced_scaling}
\end{equation}where $c$ and $d$ are different exponents than those in Eq.~\eqref{scaling}. Together with Eq.~\eqref{discussion_invariant_saffman}, this scaling leads to a different constraint on the decay,
\begin{equation}
    \psi^c L_\psi^d\,  \ldots \sim \const, \label{constraint_saffman}
\end{equation}which is independent of the fractional imbalance $\sigma$. 

We note that an eventual transition from the balanced regime  [Eqs.~\eqref{fractional_constraint} and~\eqref{constraint_saffman}] to the imbalanced regime [Eq.~\eqref{constraint}] is a general consequence of these results, because, together, Eqs.~\eqref{fractional_constraint} and~\eqref{discussion_invariant_saffman} imply
\begin{equation}
    \sigma \propto L_f^{3/2}. \label{growth_in_sigma}
\end{equation}A necessary and sufficient condition for the fractional imbalance to grow is, therefore, that the scale $L_f$ should increase with time. This tends to be the case for any realistic decay problem, even in the absence of inverse transfer of energy in $k$-space, because small-scale structures generally dissipate faster than large-scale ones.

The correspondence of Saffman-type invariants to the small-$k$ asymptotics of spectra, as explained for $I_H$ in Section~\ref{sec:formal}, is also a general property: the invariant $I_F$ is related to the spectrum of the variance of $f(\bx)$,
\begin{equation}
    \Theta_F(k) \equiv \frac{k^2}{4\pi^2} \int \dd^3 \br\, \langle f (\bx)\, f (\bx + \br) \rangle e^{-i\bk \bcdot \br},
\end{equation}via
\begin{equation}
    \Theta_F(k\to 0) = \frac{I_{F} k^2}{4 \pi ^2} + O(k^3), \label{Theta_F_expansion}
\end{equation}provided that $\langle f (\bx)\, f (\bx + \br) \rangle<O(r^{-3})$ as $r\to\infty$, and that the system is statistically isotropic (though this statement is easily reformulated for anisotropic systems). Therefore, systems that decay while respecting the conservation of a Saffman-type invariant generally have a `permanence of large scales' principle that applies to the spectrum of the variance of the relevant conserved quantity. This principle provides a convenient way to assess the existence and conservation of candidate invariants in numerical studies (as we did in Fig.~\ref{fig:helicity_spectrum}).

To conclude this section, we note that our discussion has relied on two conditions: (\textit{i}) that the invariant, $F$, is conserved, i.e., that its rate of change is smaller than the energy decay rate, and (\textit{ii}) that it can always be considered the case that typical structures possess $F$, so that $\langle f\rangle$ and $\langle f^2 \rangle$ may be related to the sizes and correlation lengths of the dynamical fields, via Eqs.~\eqref{scaling},~\eqref{fractional_scaling} and~\eqref{balanced_scaling}. Establishing whether (\textit{i}) and (\textit{ii}) holds for any given turbulent system requires some physical idea of the decay dynamics. For example, we found in Section \ref{sec:otherinvariants} that the Saffman-type invariant corresponding to cross-helicity was not conserved in our numerical simulations. Nonetheless, different dynamical processes can result in invariants being dissipated at different rates, and indeed there are plausible reasons to argue that cross-helicity (or its Saffman-type-invariant counterpart) might be conserved by the decaying turbulence of interacting nonlinear \Alfven~waves, as we will explain in Section~\ref{sec:meanfield}.

Dynamical processes may also govern whether structures possess the relevant invariants, condition (\textit{ii}). For example, we have argued that in decaying MHD turbulence, non-helical magnetic structures tend to relax to zero energy (as is consistent with J.B.~Taylor relaxation), so that at any given time, individual magnetic structures are maximally helical.

As a final general remark, we point out that it is possible for nonlinear structures to possess more than one invariant. If the constraints implied by the conservation of these invariants are not mutually exclusive, then they must be satisfied simultaneously. If, however, these conditions are contradictory, then it will be necessary for some of them to be broken. In that case, the decay cannot take place on the characteristic nonlinear timescale (``eddy-turnover time''), but must instead take place on the timescale on which the weaker constraint (in the sense of quality of conservation) can be broken. This is precisely the situation in magnetically dominated turbulence: magnetic helicity is not the only topological invariant associated with the magnetic field~\cite{Moffatt85}, and in principle all higher-order topological invariants might impose constraints on the decay. Conserving all topological invariants is not consistent with a decay in the magnetic energy, however, because it implies $B\sim \const$. Therefore, the decay can only proceed on the timescale on which the higher-order topological constraints may be violated, i.e., the reconnection timescale~\cite{Ruzmaikin94}. This is illustrative of a general rule that stronger, consistent constraints set the scalings between the integral scales and energies that control the decay, while inconsistent constraints set the decay timescale.

\subsection{When the Saffman helicity invariant fails\label{sec:Saffman_fails}}

Let us now apply the insights of Section~\ref{sec:general_principles} to determine the conditions under which the conservation of the Saffman helicity invariant does not impose a constraint on the decay of isotropic, non-helical MHD turbulence.

Naturally, this will be the case if it is zero, which is possible if each magnetic structure is \emph{individually} non-helical, i.e., $h$ is pointwise zero, so condition (\textit{ii}) in Section~\ref{sec:general_principles} is violated. An example would be an ensemble of untwisted, unlinked magnetic tori, such as arise in the final stage of process (b) shown in Fig.~\eqref{fig:cartoon}~\footnote{We are grateful to K.~Subramanian for pointing out this example to us.}. Such a configuration would not be prevented by the invariance of its magnetic topology from relaxing quickly under ideal dynamics, transferring magnetic energy to kinetic. The evolution should then be constrained by invariants pertaining to the velocity field, such as the Loitsyansky integral or cross-helicity --- plausibly, the latter may be conserved in a net [cf. Eq.~\eqref{discussion_invariant}] or fluctuating [cf. Eq.~\eqref{discussion_invariant_saffman}] sense in turbulence with $U\sim B$ and similar integral scales for both fields, as we shall explain in Sections~\ref{sec:meanfield} and \ref{sec:UsimB}.


While we thus acknowledge the possibility that $I_H$ can be zero, we view such a field configuration as rather artificial. In any realistic physical scenario, it seems likely that some non-zero fraction of magnetic structures should possess helicity, whether due to twists or to linkages, so $I_H\neq 0$. Even if this fraction is not $1$, we expect that individually non-helical structures should relax on ideal timescales (cf. Fig.~\ref{fig:cartoon}), and the evolution of those that remain should then be constrained by conservation of $I_H$ (also see \cite{Servidio08} for a study of the tendency of MHD turbulence to form local helical patches).

The other scenario in which conservation of $I_H$ will not impose a constraint on the decay is if it diverges, which it will do if the magnetic-helicity correlation function $\langle h(\bx)h(\bx+\br)\rangle$ decays as $r^{-3}$ or slower as ${r\to\infty}$. Physically, $I_H$ diverges when our assumption of `localised magnetic structures' fails. This should be expected if
\begin{equation}
    I_{\bB}\equiv \int\dd^3 \br \, \langle \bB(\bx)\bcdot\bB(\bx+\br) \rangle\label{I_B}
\end{equation}is non-zero, which, as we saw in Eq.~\eqref{EMexpansion}, corresponds to $\mcE_M(k\to0) \propto k^2$. Such a turbulence has long-range (longitudinal) correlations in the magnetic field that decay as $r^{-3}$ as $r\to \infty$ (the reason for this is the same as the one for which long-range correlations in the velocity field are required for $I_{\bP}\neq 0$ --- we refer the interested reader to \cite{Davidson13}). Magnetic-helicity correlations should be expected to decay at least as slowly as this ($\bA$ being formally an integral of $\bB$), in which case $I_H$ will diverge.

So, what principle governs the decay of non-helical turbulence of this sort? We suggest that it should be constrained by the conservation of fluctuations in the magnetic flux, under the formalism described in Section~\ref{sec:general_principles}.
In this case, $I_{\bB}$ itself is the relevant invariant, which we might call the `Saffman flux invariant' [cf.~Eq.~\eqref{discussion_invariant_saffman}], as
\begin{equation}
    I_{\bB} = \lim_{V\to\infty}\frac{1}{V}\langle |\boldsymbol{\Phi}_V|^2 \rangle,
\end{equation}where $\boldsymbol{\Phi}_V = \int_V \dd^3 \bx \,\bB$ is the total flux contained within the control volume $V$. In light of Eq.~\eqref{EMexpansion}, we note that non-helical turbulence with $\mcE_M(k\to 0)\propto k^2$ should not have an inverse energy transfer; its energy spectrum should instead obey a `permanence of the large scales' principle. This prediction is consistent with the results of the extensive parameter study by \cite{ReppinBanerjee17}. In decay from a magnetically dominated state,
\begin{equation}
    I_{\bB}\sim B^2 L^3\sim \const
\end{equation}
(i.e., $\alpha = 2/3$) corresponds to a decay law of 
\begin{equation}
    E_M \propto t^{-6/5}
\end{equation}if the decay occurs on ideal timescales [see Eq.~\eqref{ideal_law}]. This is, inevitably, the same as Saffman's law for the decay of kinetic energy in hydrodynamic turbulence with ${\mcE(k\to 0)\propto k^2}$ \cite{Saffman67}. Assuming that fast-reconnection outflows are the dominant motions, Eq.~\eqref{EK_sim_dL_EM} with $\delta/L\sim \const$ gives $E_K\propto E_M$ (though $E_K\ll E_M$), so $E_K$ obeys the same law as $E_M$. If reconnection occurs slowly, in the Sweet-Parker regime, then Eqs.~\eqref{reconnecting_plaw} and~\eqref{kinetic_energy_decay} give
\begin{equation}
    E_M \propto t^{-4/3},\quad E_K \propto t^{-5/3},
\end{equation}for Laplacian ($n=2$) dissipation, where the latter law again assumes the dominant motions to be the reconnection outflows.

These are interesting predictions to test. However, while it is possible to initialise turbulence in a state with $I_{\bB}\neq0$, the requirement of strong long-range correlations in $\bB$ would appear to make it somewhat artificial. For example, in cosmological contexts, causality constraints are expected to rule out spectra shallower than the `Batchelor' $\mcE_M(k\to 0)\propto k^4$ \cite{DurrerCaprini03, Brandenburg15, Brandenburg17,ReppinBanerjee17}, which, as explained above, corresponds to localised magnetic structures (i.e., rapidly decaying correlations).

\subsection{Decay of MHD turbulence in the presence of strong mean field \label{sec:meanfield}}

In this work, we have so far restricted attention to the decay of isotropic MHD turbulence, i.e., MHD turbulence without a mean magnetic field. The case of turbulence with a strong mean field is fundamentally different, because then the magnetic helicity is not a conserved quantity. Formally, this is because the mean field `sticks out' of any volume for which one might choose to compute the magnetic helicity. Intuitively, also, this is a different situation compared to isotropic turbulence, because the constraints imposed by topology are much reduced. For example, purely magnetic structures need not persist until they are able to reconnect with each other, instead they can relax by decomposing themselves into \Alfven~waves travelling in opposite directions along the mean field.

Aside from energy, MHD turbulence with a strong mean field (described by the `reduced MHD' equations~\cite{Strauss76, KadomtsevPogutse74}) has only one conserved quantity related to the presence of the magnetic field, the cross-helicity,
\begin{equation}
    H_c = \int \dd^3 \br \, \bu_{\perp} \bcdot \, \bb_{\perp},
\end{equation}where $\bu_\perp$ is the fluid velocity perpendicular to the mean magnetic field, and $\bb_\perp$ is the magnetic-field perturbation. Like magnetic helicity, the cross-helicity is sign-indefinite. In simulations of driven MHD turbulence with a strong mean field, a tendency to develop patches of strong local cross-helicity is observed, even in so-called balanced turbulence where the volume-averaged cross-helicity density is zero~\cite{PerezBoldyrev09, StriblingMatthaeus91,Ting86,Servidio08,Matthaeus08}. The reason for this is that structures of large cross-helicity are also structures containing strong imbalance in the sizes of the two \Elsasser~fields, $\bZ^{\pm}=\bu_{\perp} \pm \bb_{\perp}$. Importantly, individual \Elsasser~fields each represent exact nonlinear solutions to the MHD equations. Nonlinearity, and hence turbulent decay, can, therefore, only be present where both fields are present. Since $\bu_\perp \bcdot \bb_\perp = (|\bZ^+|^2 - |\bZ^-|^2)/4$, a small cross-helicity density indicates balance between the \Elsasser~fields, and, therefore, large nonlinearity. Such structures are prone to turbulent decay. In contrast, structures with strong cross-helicity of either sign have reduced nonlinearity and, therefore, are more immune to turbulent decay.

Note, however, that these considerations need not apply to isotropic, magnetically dominated MHD turbulence decaying via reconnection, because the possession of cross-helicity does not afford immunity to reconnection. For balanced, Reduced-MHD turbulence, however, they motivate us to conjecture that the decay might be controlled by the `Saffman cross-helicity invariant'
\begin{equation}
    I_{H_c} =  \int \dd^3 \br \langle h_c(\bx) h_c(\bx+\br)\rangle, \label{saffman_crosshelicity_invariant}
\end{equation}where $h_c = \bu_{\perp}\bcdot \bb_\perp$ \footnote{MHD relaxation subject to net-cross-helicity conservation has been considered as part of the general selective-decay framework by \cite{Montgomery78, MatthaeusMontgomery80, Ting86, Hossain95, Wan12}, and has been termed `dynamical alignment' (though this should not be confused with the use of the same term to describe the alignment of Els{\"{a}}sser fields in the inertial range of MHD turbulence \cite{PerezBoldyrev09, Mallet15, Schekochihin20}), owing to the tendency for a fractionally cross-helical state to reach the maximally cross-helical state (and hence, cease decaying) \cite{Oughton94, MaronGoldreich01,Cho02, Chen11}, which agrees with the general theory presented in Section \ref{sec:general_principles}. In the case of small or zero total cross-helicity, however, such conclusions do not apply, and one must consider the \emph{Saffman} cross-helicity invariant, Eq.~\eqref{saffman_crosshelicity_invariant}, as the relevant one.}. We note that, like $I_{H}$, $I_{H_c}$ is an example of an invariant that depends on a fourth-order correlation function. The relevance of fourth-order correlators to distinguishing between different species of decaying MHD turbulence has previously been suggested by \cite{Wan12}, inspired by the numerical results of~\cite{Lee10}. 

One might expect $I_{H_c}$ to be finite and conserved by precisely the same arguments as we presented for $I_H$, the Saffman helicity invariant, in Section~\ref{sec:non-helical}. By a random-walk argument analogous to the one in Section~\ref{sec:formal}, we have 
\begin{equation}
    I_{H_c} \sim b_\perp^2 u_\perp ^2 l_\perp^2 l_\parallel,
\end{equation}where $l_\parallel$ and $l_\perp$ are the characteristic lengthscales parallel and perpendicular to the mean field, respectively. For Alfv\'{e}nic motions, $b_\perp \sim u_\perp$. Note that this scaling is on much firmer ground in the mean-field case than the isotropic case, because of the absence of topological constraints associated with helicity conservation. The parallel and perpendicular length scales can be related to each other by the conjecture of critical balance \cite{Zhou20}, which states that $b_\perp / l_\perp \sim B_0 / l_\parallel$, and is a cornerstone of the theory of strong MHD turbulence~\cite{GS95, NazarenkoSchekochihin11, Stawarz12}. Critical balance is essentially a statement of causality: the characteristic parallel length scale of an eddy cannot be longer than the distance travelled by an \Alfven~wave in one nonlinear timescale. It has been confirmed numerically to great precision in driven RMHD turbulence~\cite{Mallet15}, and appears to be satisfied in decaying turbulence too~\cite{Zhou20}. 

Putting these scalings together, we find that $I_{H_c} \propto b_\perp^3 l_\perp ^3$. If $I_{H_c}$ is indeed conserved in decaying MHD turbulence, this implies
\begin{equation}
    b_\perp l_\perp \sim \const.
\end{equation}Then, since the energies of the \Elsasser~ fields are comparable in balanced turbulence, the turbulent decay would likely be governed by
\begin{equation}
     \frac{\dd E_{Z^{\pm}}}{\dd t} \sim  - \frac{ Z^{\mp} E_{Z^{\pm}}}{l_\perp} \sim - {E_{Z^{\pm}}^2}.
\end{equation}This results in a decay of both magnetic and kinetic energy
\begin{equation}
    E_M\sim E_K \sim t^{-1}.\label{meanfieldlaw}
\end{equation}Intriguingly, this decay law has indeed been observed in simulations of decaying, balanced RMHD turbulence, though was rationalised differently, by assuming local effective conservation of anastrophy, which also implies Eq.~\eqref{meanfieldlaw}, as in two-dimensional MHD turbulence~\cite{Zhou20} (see Appendix~\ref{sec:2d}).

If indeed the decay of MHD turbulence in the presence of a strong mean field conserves cross-helicity, any initial imbalance will eventually lead to a final state with maximal cross-helicity, i.e., a pure \Elsasser~state, according to the general argument presented in Section~\ref{sec:general_principles} [see Eq.~\eqref{growth_in_sigma}]. Such a state will not decay, since it is an exact solution of the non-linear RMHD equations. Such behaviour has indeed been reported in numerical studies \cite{MaronGoldreich01, Oughton94, Cho02, Chen11}.

\subsection{Decay of isotropic MHD turbulence from an initial state with $U\sim B$\label{sec:UsimB}}

Another problem to which the formalism developed here may be usefully applied is to the decay of isotropic MHD turbulence from an initial state with $U\sim B$, as opposed to the $B\gg U$ that we have so far considered in this work (apart from in Section~\ref{sec:meanfield}). Such a state is the natural final state of the MHD dynamo (see, e.g., \cite{Rincon19, Schekochihin20} for reviews). In such a case, we conjecture that the simultaneous conservation of magnetic helicity \textit{and} cross-helicity might be respected by the decay, as the constraints they imply are not mutually exclusive. Such a prospect has been considered by \cite{Montgomery78, MatthaeusMontgomery80, Ting86, Hossain95, Wan12} for net-cross-helical initial states --- here, however, we shall impose cross-helicity conservation via the Saffman-type-invariant formalism developed in Section \ref{sec:general_principles}, reflecting the fact that no strong net cross-helicity is generically present in MHD turbulence without a mean magnetic field. Naturally, checking the circumstances under which conservation of Eq.~\eqref{saffman_crosshelicity_invariant} (with $h_c = \bu \bcdot \bB$ in the isotropic case) is valid will require a detailed numerical study, which is left for future work. However, the consequences of this conjecture merit discussion here, because they do appear to be consistent with already-existing numerical studies, and the argument demonstrates a general principle of simultaneous conservation of multiple invariants.

Let us consider a system that, as a result of dynamo or some other process, has reached equipartition between magnetic and kinetic energy, $U\sim B$, with the same integral scale $L$.

\subsubsection{Helical magnetic field}

First, let us assume that the magnetic field is helical, but that there is no predominance of either sign of cross-helicity. Then the conjecture of simulataneous conservation of magnetic helicity and cross-helicity (the latter as a Saffman-type invariant) implies $B^2 L \sim \const$ [Eq.~\eqref{B2L}] and $B^2 U^2 L^3 \sim \const$, respectively. Together, these conditions imply $U\sim B^2$, or 
\begin{equation}
    E_K \sim E_M^2 \label{helicalUsimB}
\end{equation}precisely the condition found numerically by \cite{BiskampMuller99, BiskampMuller00}, though without theoretical justification. They conjectured that the decay should take place on the timescale associated with the $\bu \bcdot \nabla \bu$ nonlinearity in the MHD equations, i.e., $L/U$, which is consistent with the idea that the timescale associated with the magnetic nonlinearity $\bB \bcdot \nabla \bB$ is effectively lengthened by topological constraints on the magnetic field. It is readily verified that such a decay leads to the power laws
\begin{equation}
    E_K\sim t^{-1},\quad E_M\sim t^{-1/2}, \label{plaws_simulataneous}
\end{equation}as found numerically by \cite{BiskampMuller99, BiskampMuller00, BereraLinkmann14}.

The more rapid decay in kinetic energy will result in a state with $U\ll B$. The system should then decay in the strong-field regime described in Section~\ref{sec:meanfield}. Denoting the small perturbations to the newly established strong magnetic field by $\delta B \sim U$, we should, according to Eq.~\eqref{meanfieldlaw}, then have $\delta B^2 \sim U^2 \sim t^{-1}$, assuming the decay is critically balanced. Meanwhile, $B^2$ should decay according to the reconnection-controlled law, either $t^{-2/3}$ for fast reconnection, or $B^2 \sim t^{2n/(5n-3)}$ [Eq.~\eqref{generaln_helical}] for Sweet-Parker reconnection. For the $n=4$ simulations employed by \cite{BiskampMuller99, BiskampMuller00}, this implies
\begin{equation}
    E_K\sim t^{-1},\quad E_M\sim t^{-8/17}, \label{plaws_simulataneous2}
\end{equation}which will persist until the Sweet-Parker outflows dominate the kinetic energy, at which point the kinetic-energy law should change to the one given by Eq.~\eqref{kinetic_energy_decay}, which, independently of the type of reconnection, is always slower than $t^{-1}$. 

The decay laws given by Eq.~\eqref{plaws_simulataneous2} are very close to those of Eq.~\eqref{plaws_simulataneous}, and, therefore, either might explain the laws found numerically by \cite{BiskampMuller99, BiskampMuller00}. Regardless, a more rapid decay of the kinetic than magnetic energy appears to be robust, as does the corresponding establishment of magnetically dominated state. Indeed, a magnetically dominated final state was observed in the simulations of \cite{Brandenburg19}, despite being initialised with $U\gg B$, with decay laws similar to Eq.~\eqref{plaws_simulataneous} in the transient $U\sim B$ regime.

\subsubsection{Non-helical magnetic field \label{sec:nonhelical_speculation}}

Alternatively, let us consider the case of a non-helical magnetic field, initially in equipartition with the kinetic energy. Then, instead of the condition $B^2 L\sim \const$, we have $B^4 L^5 \sim \const$ [Eq.~\eqref{B45L}]. Together with the constraint implied by the conservation of the Saffman cross-helicity invariant, $B^2 U^2 L^3 \sim \const$, this implies $U\sim B^{1/5}$, or 
\begin{equation}
    E_K \sim E_M^{1/5}. \label{nonhelicalUsimB}
\end{equation}Unlike Eq.~\eqref{helicalUsimB}, Eq.~\eqref{nonhelicalUsimB} implies a much faster decay of the magnetic energy than the kinetic energy. However, this decay will be short lived, because the magnetic energy can be maintained at a small, but finite, fraction of the kinetic energy by dynamo. Nonetheless, the magnetic field will always remain just below dynamical strength, because if it were to grow to dynamical levels, cross- and magnetic helicity conservation would force it to decay rapidly.

In the absence of a dynamical-strength magnetic field, the kinetic energy should decay according to the purely hydrodynamic Kolmogorov law, $E_K\sim t^{-10/7}$ [Eq.~\eqref{t10/7}]. Because the magnetic energy is tied to a finite fraction of the kinetic energy by the competing effects of dynamo and simultaneous cross- and magnetic-helicity conservation, it must also be the case that $E_M\propto t^{-10/7}$, i.e.,
\begin{equation}
    E_M \propto (\mathrm{but }<\,) \,E_K \sim t^{-10/7}.\label{nh_speculative_law}
\end{equation}

Such a decay of $E_M$ and $E_K$ has indeed been observed in non-helical simulations (though initialised with $U\gg B$), by \cite{Bhat20}, together with evidence of dynamo action. We note, however, that it is possible that the reason $E_M$ never grew to equipartition in these simulations was, rather, that the efficiency of the dynamo was reduced in the absence of the mean-field dynamo effect associated with helical velocity fields (see \cite{Rincon19} and references therein).

The arguments presented in this section, if correct, suggest a remarkable general principle: an initially helical velocity field will, due to its tendency to grow  a helical magnetic field through mean-field dynamo action, eventually decay in a magnetically dominated state, with $E_M,\,E_K \propto t^{-2/3}$ (in the fast reconnection regime). In contrast, non-helical velocity fields will always remain in a kinetic-energy-dominated state, with $E_M,\,E_K \propto t^{-10/7}$.

\subsection{Conclusion}

As Sections~\ref{sec:meanfield} and~\ref{sec:UsimB} illustrate, the Saffman-type-invariant approach appears to be an extremely useful general tool that allows consideration of sign-indefinite invariants in the `selective decay' framework for decaying turbulence 
\footnote{Saffman-type invariants might also be usefully incorporated, as constraints, into theories of Gibbsian relaxation to absolute equilibrium states \cite{StriblingMatthaeus90, StriblingMatthaeus91}, which have been shown to be a predictor of the late-time properties of relaxing MHD systems \cite{Stawarz12}.}.
There are many different types of fluid turbulence to which the 
approach may be usefully applied --- a number of them are reviewed in \cite{Davidson13}, and there are likely to be others, especially in the large variety of plasma systems increasingly of interest in the context of various types of space or astrophysical turbulence (see, e.g., \cite{Schekochihin19, Milanese20, Meyrand21}). The invariant $I_H$, proposed in this work for incompressible magnetohydrodynamic turbulence, should remain an invariant even in the more complex cases of compressible~\cite{MacLow98, Stone98}, relativistic~\cite{Zrake14} and/or kinetic dynamics~\cite{Nalewajko16, Yuan16, Lyutikov17a, Nalewajko18}, since magnetic helicity remains a conserved quantity in such contexts. This means that the invariance of $I_H$  (and the physical principle of conservation of local magnetic-helicity fluctuations from which it follows) should provide a constraint on decaying turbulence in a wide variety of magnetised astrophysical systems (provided the dominant motions occur at scales much smaller than the system size). Possible applications include turbulence in star-forming molecular clouds~\cite{MacLow98, Gao15} and galaxy clusters~\cite{Subramanian16, Sur19}; the generation of seed fields for galactic dynamo \cite{Zhou20}; and the evolution of primordial magnetic fields in the early Universe \cite{BanerjeeJedamzik04, DurrerNeronov13, Subramanian16}. With regard to the latter, we note that the non-helical decay laws that we have derived in this work [see Eq.~\ref{nonhelical_laws}] are consistent with observational constraints on magnetic fields in cosmic voids, whereas previously accepted models are not~\cite{WagstaffBanerjee16}. This point will be addressed specifically in a future publication~\cite{HoskingSchekochihin21primordial}.

In those physical systems where frozen-in magnetic fields are dominant actors in the dynamics, magnetic reconnection is likely to be the key physical process whereby the decay occurs. Marrying this insight to the constraints imposed by invariants appears to be a winning strategy for constructing decay theories, as it has proven to be in the MHD turbulence regimes that we have considered above.

\begin{acknowledgments}

We are grateful to P.~A. Davidson, N.~F. Loureiro, J.~C. Miller, F. Rincon, D. Uzdensky, and M. Zhou for discussions of this work, and to G. Lesur for providing us with an up-to-date version of the Snoopy code. D.N.H. was supported by a UK STFC studentship. The work of A.A.S. was supported in part by the UK EPSRC grant EP/R034737/1. This work used the ARCHER UK National Supercomputing Service (http://www.archer.ac.uk).

\end{acknowledgments}

\appendix

\section{\label{sec:2d}Decay of two-dimensional turbulence}\FloatBarrier

\begin{figure*}
\includegraphics[width=\textwidth]{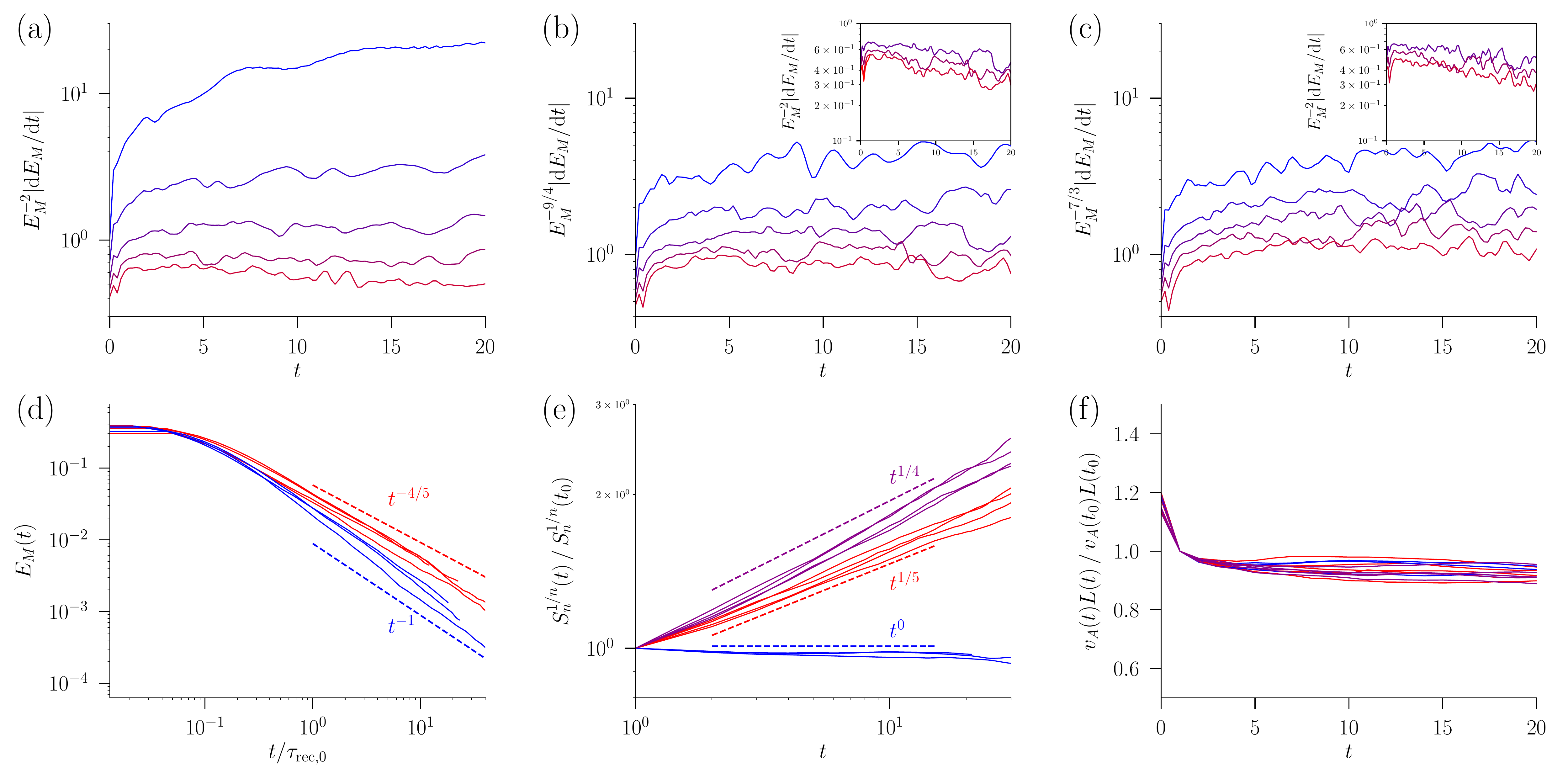}
\caption{\label{fig:2D_mega} Evolution of average quantities during our two-dimensional simulations. (a) Normalised energy-decay rate for $n=2$, (b) $n=4$, (c) $n=6$. In each case, a flat profile indicates agreement with the expected decay power laws, which are $t^{-1}$, $t^{-4/5}$ and $t^{-3/4}$, respectively [Eq.~\eqref{2d_rec_plaw}]. Blue $\to$ red indicates increasing~$S_n^{1/n}$. Insets show the scaling that would be expected if the ideal law $E_M\sim t^{-1}$ were satisfied [Eq.~\eqref{2d_ideal_plaw}]. (d) Magnetic energy evolution for $n=2$ (blue) and $n=4$ (red) plotted against time renormalised to the reconnection timescale at $t=1$, as in~\cite{Zhou19}. (e) Growth of the hyper-Lundquist number with time, for $n=2$ (blue), $n=4$ (red), and $n=6$ (magenta) simulations. Dashed lines indicate the expected scalings. (f) Constancy of $BL$ with time. N.B. the two $n=2$ runs with smallest $S_n^{1/n}$ are not plotted in (d), (e), (f) as they do not exhibit constant $BL$ (see Table \ref{tab:table}).}
\end{figure*}

In this appendix, we review the problem of decaying two-dimensional MHD turbulence, which, like its three-dimensional counterpart, respects the conservation of an invariant associated with the topology of the magnetic field: the square of the magnetic vector potential,  $\langle A^2 \rangle = \langle A_z^2 \rangle$ \cite{MatthaeusMontgomery80,Hatori84,Ting86,BiskampWelter89}, sometimes called `anastrophy'. In two dimensions, anastrophy is well defined (i.e., not gauge dependent), and evolves according to
\begin{align}
    \left| \frac{\mathrm{d}}{\mathrm{d} t} \int_V \dd^2 \br A^2 \right| & = 2 \eta_n \left| \int_V \dd^2 \br \bA \bcdot \nabla^n \bA \right| \nonumber \\
    & = 2 \eta_n \left| \int_V \dd^2 \br \bB \bcdot \nabla^{n-2} \bB \right| \nonumber \\
    & \sim \frac{\dd E_M}{\dd t} {\delta_\eta}^2.\label{MHD_2D_Az_cons}
\end{align}where $\dd E_M/\dd t = \eta_n\int\mathrm{d}^2\boldsymbol{r}\,\bB \bcdot \nabla^n \bB$ is the rate of magnetic-energy decay due to Ohmic heating. Eq.~\eqref{MHD_2D_Az_cons} implies that 
\begin{equation}
    \frac{\dd \log \int_V \dd^2 \br\, A^2}{\dd t} \sim  \frac{{\delta_{\eta}}^2}{L^2} \frac{\dd \log E_M}{\dd t},\label{MHD_2D_Az_cons2}
\end{equation}which is an even slower rate of change than the one we found for the magnetic helicity in three dimensions, Eq.~\eqref{scaling_for_helicity_decay}. Therefore, like helicity, anastrophy in two dimensions should be conserved as the turbulence decays, for $\eta_n \to 0^+$. Physically, anastrophy conservation is related to the conservation of in-plane magnetic flux \cite{Zhou19}. Unlike helicity, though, the anastrophy is manifestly positive definite, so there is only one decay regime, and no Saffman-type invariant.

The conservation of anastrophy implies 
\begin{equation}
    BL\sim \const, \label{BL}
\end{equation}i.e., $\alpha = 1$ in Eq.~\eqref{BalphaL}. According to Eqs.~\eqref{ideal_law} and~\eqref{reconnecting_plaw}, this implies a power law decay of the magnetic energy 
\begin{equation}
    E_M\sim t^{-1}\label{2d_ideal_plaw},
\end{equation}if the decay proceeds on the `ideal' $L/B$ timescale, and
\begin{equation}
    E_M\sim t^{-2n/(3 n - 2)},\label{2d_rec_plaw}
\end{equation}for a decay proceeding on the Sweet-Parker timescale, $(L/B)\, S_n^{1/n}$. Coincidentally, these laws are the same for $n=2$. Remarkably, the Sweet-Parker scaling for the kinetic-energy decay also turns out to be 
\begin{equation}
    E_K \sim \frac{\delta}{L} E_M \sim E_M^{(3n-2)/2n} \sim t^{-1}\label{rec_KE_plaw_2d},
\end{equation}which is independent of $n$. These results mean that a decay on ideal timescales produces indistinguishable power laws ($E_K \sim E_M \sim t^{-1}$) to a decay on the $n=2$ Sweet-Parker timescale. It is perhaps for this reason that over thirty years separates the derivation of the ideal decay law~\cite{Hatori84,BiskampWelter89} and the suggestion of a decay controlled by Sweet-Parker reconnection by \cite{Zhou19}. The picture presented in \cite{Zhou19} is analogous to (and has inspired) the `cartoon' model that we proposed in Section \ref{sec:non-helical}, although we observe that, as for the cartoon model in our study, the formulation based on the conservation of integral invariants, with decay proceeding on the reconnection timescale, is more general, as it does not require that pairwise mergers between structures of equal anastrophy (or helicity) be the only allowed dynamical process.

As a consequence of the degeneracy in power laws, it was confirmed in \cite{Zhou19} that the Sweet-Parker timescale governed the decay in their simulations by showing that their evolution curves collapsed onto each other when time was renormalised by the initial Sweet-Parker reconnection timescale. An alternative method by which the Sweet-Parker-controlled decay can be established is via the use of hyper-dissipation, in the same manner as we have done in the main part of this work, thanks to hyper-dissipation lifting the power-law degeneracy. For example, from Eq.~\eqref{2d_rec_plaw}, the magnetic-energy-decay power laws are $t^{-4/5}$ and $t^{-3/4}$ for $n=4$ and $n=6$, respectively. 




In Fig.~\ref{fig:2D_mega}(a-c), we show the evolution of the magnetic-energy decay rate in our two-dimensional simulations (see Table~\ref{tab:table} in Appendix~\ref{sec:app_plaws} for details), normalised by the power of the energy to which it is proportional in the reconnection-based theory [i.e., the power of $B^2$ on the right hand side of Eq.~\eqref{reconnecting_DE}, with $\alpha =1$]. As explained in Appendix \ref{sec:app_plaws} (also, see \cite{BiskampMuller99}), such plots are preferable to more conventional plots of $\log E_M$ against $\log t$, because they give an unbiased estimate of the decay law. On these plots, horizontal curves indicate agreement with theoretical expectations. As the resistivity decreases, and hence the Lundquist number increases, Figs.~\ref{fig:2D_mega}(a-c) show increasingly horizontal curves, in agreement with Eq.~\eqref{2d_rec_plaw}. The insets to these figures show the scaling that would be expected if the decay proceeded via ideal motions~\cite{Hatori84,BiskampWelter89}. In both cases, we find clearly decreasing curves for the largest Lundquist numbers tested, so these results are inconsistent with the `ideal' law, Eq.~\eqref{2d_ideal_plaw}.

In Fig.~\ref{fig:2D_mega}(d) we show $E_M$ against $t/\tau_{\mathrm{rec, 0}}$, where $\tau_{\mathrm{rec, 0}}$ is the Sweet-Parker timescale $S^{1/2}L/B$ at the start of the self-similar decay period, which we take to occur at $t=1$ in all cases for the purposes of this calculation. While such plots are not well-suited to an accurate determination of the decay law, they do show a clear difference in behaviour between the case of Laplacian dissipation ($n=2$, blue) and hyper-dissipation ($n=4$, red). As previously mentioned, the collapse of the decay curves onto each other under such a normalization was presented as evidence for a reconnection-controlled decay in two-dimensions by ~\cite{Zhou19}. Fig.~\ref{fig:2D_mega}(d) shows that the same behaviour occurs in the $n=4$ hyper-dissipative case, and we find the same in the $n=6$ case (not shown), so we agree entirely with~\cite{Zhou19} that the decay is controlled by reconnection.

The remaining panels of Fig.~\ref{fig:2D_mega} show other relevant quantities besides the magnetic energy. In Fig.~\ref{fig:2D_mega}(e), we show the evolution of the hyper-Lundquist number, which is in excellent agreement with theoretical expectations based on Eqs.~\eqref{BL} and~\eqref{2d_rec_plaw}. Interestingly, the hyper-Lundquist number is expected to grow when $n>2$, so even if the simulation starts in the Sweet-Parker reconnection regime, it can ultimately transition to the plasmoid-dominated regime. We reiterate that in such a regime, which is formidable to simulate, the reconnection timescale becomes proportional to the ideal timescale (though longer by a factor of $10^{2}$ \cite{Uzdensky10}), and then we expect a transition to the `ideal' $t^{-1}$ decay [Eq.~\eqref{2d_ideal_plaw}]. Finally, Fig.~\ref{fig:2D_mega}(f) confirms that $BL$ is indeed constant during the decay, as demanded by the conservation of anastrophy.

\begin{figure}
\includegraphics[width=\columnwidth]{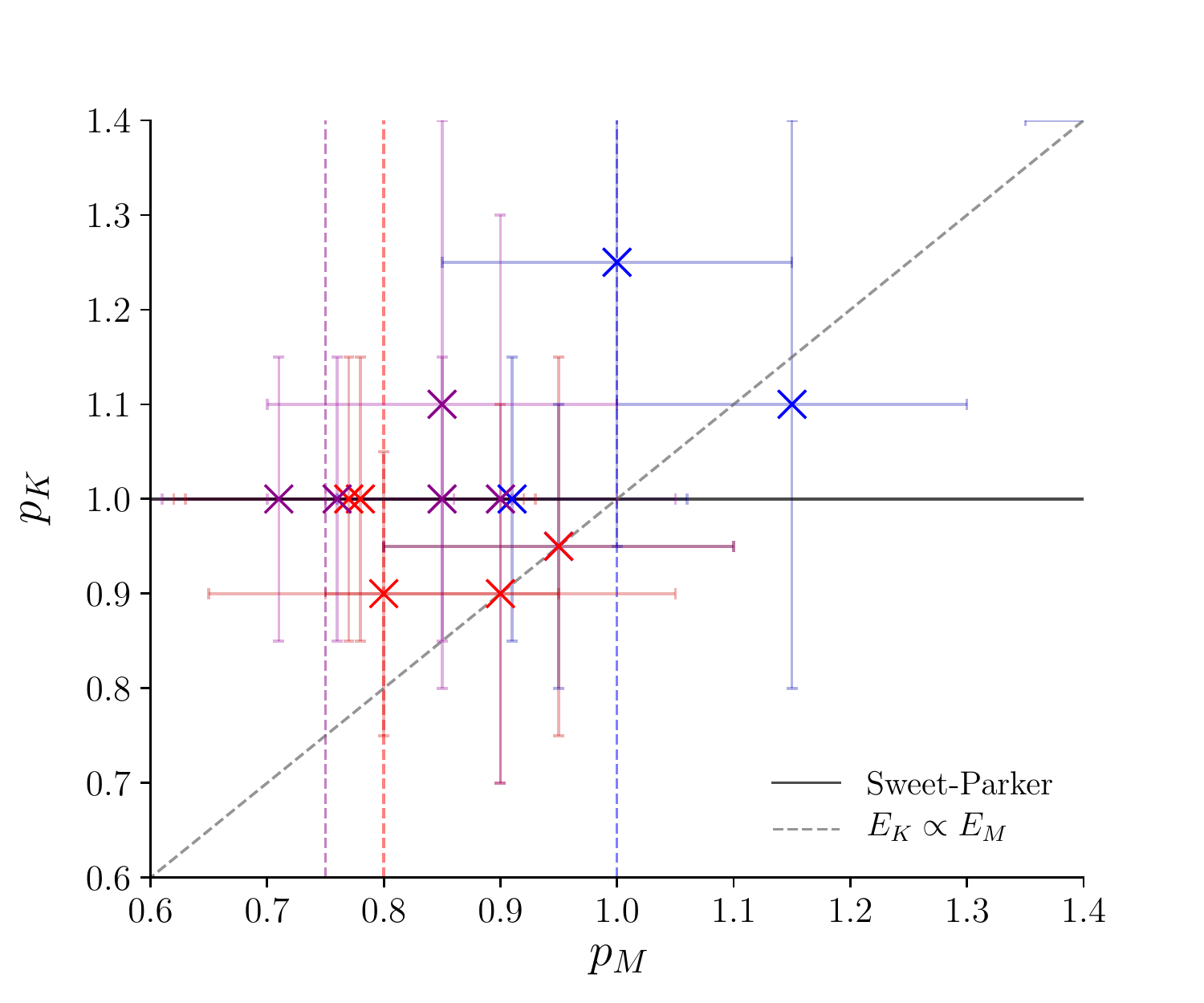}
\caption{\label{fig:EK_2d} Relation between the measured magnetic- and kinetic-energy decay exponents in our two-dimensional simulations. Eq.~\eqref{rec_KE_plaw_2d} predicts that all simulations should have $p_K=1$, independently of $p_M$. While many of the simulations do fall close to this line, the evidence for $p_K=1$ over $p_K=p_M$ is not very strong, owing to the closeness of all powers involved, and the large amount of noise in the two-dimensional simulations (see end of Appendix~\ref{sec:2d}).}
\end{figure}

Consider now the decay of the kinetic energy. Fig.~\ref{fig:EK_2d} shows plots of our best numerical estimates of $p_K$ vs.~$p_M$. While we note that most of our runs do fall close to the $p_K=1$ line, as predicted by Eq.~\eqref{rec_KE_plaw_2d}, the numerical evidence for $p_K=1$ over $p_K=p_M$ is not very strong, due to the similarity between the decay laws involved. Another factor that makes this comparison difficult is the high level of noise in the decay curves (see Fig.~\ref{fig:2d_superfigure}), which is not present in three dimensions, and arises because of the greatly reduced number of magnetic structures in the two-dimensional simulations. Because we initialise both types of simulation with the magnetic-spectral-energy density peaked at the same wavenumber, $k_p\simeq 33$ (see Appendix \ref{sec:app_plaws}), there are initially $\sim 33$ times fewer structures in our two-dimensional runs.

\section{Alternative proof of the conservation of the Saffman helicity invariant \label{sec:app_invariance_proof}}

In this appendix, we present an alternative proof of the invariance of $I_H$, that follows directly from the MHD induction equation. As in Section~\ref{sec:formal}, we shall rely on the assumption of rapidly decaying spatial correlations.

Restricting to the physical case of Laplacian ($n=2$) dissipation, for simplicity, the MHD induction equation reads
\begin{equation}
    \frac{\p \bB}{\p t} = \bnabla \times (\bu \times \bB - \eta \bnabla \times \bB).\label{induction}
\end{equation}Uncurling this equation, we have
\begin{equation}
    \frac{\p \bA}{\p t} = \bu \times \bB - \eta \bnabla \times \bB + \bnabla \chi,\label{dA/dt}
\end{equation}where $\chi$ is an arbitrary scalar function. From Eqs.~\eqref{induction} and \eqref{dA/dt}, it follows that
\begin{equation}
    \frac{\p h}{\p t} + \bnabla \bcdot \boldsymbol{F} = - 2\eta \bB \bcdot (\bnabla \times \bB)\label{dh/dt}
\end{equation}where, as elsewhere, $h = \bA \bcdot \bB$ is the magnetic helicity density, while
\begin{equation}
    \boldsymbol{F} = \bu (\bA\bcdot \bB) - \bB (\bA\bcdot \bu) - \chi \bB -\eta \bA \times (\bnabla\times\bB)
\end{equation}is the `magnetic-helicity flux'.
As argued in Section~\ref{sec:helicity_conservation}, the resistive helicity-dissipation term on the right-hand side of Eq.~\eqref{dh/dt} is small --- its size is $\eta B^2/\delta_{\eta}$, giving a helicity-dissipation timescale of $\sim L \delta_{\eta}/\eta$, which is long compared to the magnetic-energy dissipation timescale, $ \delta_{\eta}^2/\eta$. Dropping it, Eq.~\eqref{dh/dt} is a continuity equation for magnetic helicity. Integrating over space, we obtain its conservation law.

Alternatively, we may use Eq.~\eqref{dh/dt}, applied at $\bx$ and $\bx+\br$, to compute the evolution of the two-point helicity correlation function. After taking an ensemble average, and using statistical isotropy and homogeneity, we have
\begin{equation}
    \frac{\p}{\p t} \langle h(\bx)h(\bx+\br) \rangle + 2 \frac{\p}{\p \br} \bcdot \langle h(\bx) \boldsymbol{F}(\bx + \br)\rangle=0.\label{dhh/dt}
\end{equation}Integrating over $\br$, assuming ${\langle h(\bx) \boldsymbol{F}(\bx + \br)\rangle < O(r^{-3})}$ as $r\to \infty$, Eq.~\eqref{dhh/dt} gives
\begin{equation}
    \frac{\dd I_H}{\dd t}=0.
\end{equation}

The conservation of general Saffman-type invariants (see Section~\ref{sec:general_principles}) may be shown from their corresponding continuity equations in precisely the same manner.

\section{Random-walk scalings for linear and angular momentum \label{sec:LoitSaffDiscuss}}

In this appendix, we provide simple arguments for the scalings of $\langle|\bP_V|^2\rangle$ and $\langle|\bL_V|^2\rangle$ vs. $R$, based on the random-walk argument employed in the main text. These are, respectively, the expectation values of the squared linear and angular momenta contained within a control volume $V\sim R^3$. In the latter case, we take $V$ to be spherical, and compute the angular momentum about its centre. As noted in Section~\ref{sec:otherinvariants}, these quantities do not necessarily scale with $V$ as suggested by the na\"{i}ve random-walk estimates employed in Section~\ref{sec:formal}. However, the correct scalings may be obtained if one accounts for incompressibility, as we now demonstrate.

First, let us consider $\langle|\bP_V|^2\rangle$ for a volume $V\sim R^3$. The random-walk argument put forward in Section~\ref{sec:formal} suggests that $\langle|\bP_V|^2\rangle \propto R^3$. This is the correct scaling for Saffman turbulence, where the velocity field is initialised with long-range correlations and individual eddies can have non-vanishing linear momentum~\cite{Saffman67, Davidson13}. For $\bu$ initialised with a (longitudinal) correlation function that falls off with distance sufficiently rapidly (Batchelor turbulence), however, the correct scaling turns out to be $\langle|\bP_V|^2\rangle \propto R^2$ \cite{Saffman67, Davidson13}. We can understand this from the random-walk argument, taking into account incompressibility, as follows:
\begin{align}
    \langle |\bP_V|^2\rangle & = \left\langle\left[\int_V \dd^3 \bx \, \bu \right]^2\right\rangle = \left\langle\left[\int_S \dd \boldsymbol{S} \times \bpsi \right]^2\right\rangle\nonumber\\& \sim L^4 U^2 R^2, \label{Pscaling}
\end{align}where $\bu = \boldsymbol{\nabla} \times \bpsi$, and we have assumed that $\int_S \dd \boldsymbol{S} \times \bpsi$ will sum as a random walk (an assumption that fails if the long-range correlations in $\bu$ are strong). Therefore, the Saffman integral, 
\begin{equation}
    I_{\bP} = \lim_{V\to\infty}\frac{1}{V}\langle|\bP_V|^2\rangle\to 0
\end{equation}for such turbulence. Note also that the linear-momentum fluctuation in $V$ will not be conserved, because it is formally the same size, $\propto R$, as the net surface flux that can cause it to change. For the same reason, the conservation of the total magnetic flux $\int \dd^3 \bx \,\bB$ also does not provide a constraint on the decay of turbulence without strong long-range correlations in $\bB$ (see Section~\ref{sec:Saffman_fails} for a discussion of turbulence that does have such correlations).

Even if the Saffman integral vanishes, the local fluctuations in the linear momentum dominate $\langle|\bL_V|^2\rangle$, over the local rotation of the eddies. The reason is that structures further from the origin will contribute more angular momentum than closer structures, leading to $\langle|\bL_V|^2\rangle > O(R^3)$ (this effect also means that $\langle|\bL_V|^2\rangle$ depends on the shape of $V$, which is why it was necessary to assume $V$ to be spherical). In Saffman turbulence, where correlations are long-range and individual flow structures may have finite linear momentum, it turns out that $\langle|\bL_V|^2\rangle\sim R^5$, owing to this effect \cite{Davidson13}. This conclusion too can be obtained from a random-walk argument: the expected square angular momentum in a spherical shell of radius $r$ and width $\delta r $ satisfying $L \ll \delta r \ll r$, is
\begin{equation}
    \delta \langle |\bL|^2 \rangle \sim r^2 U_{t}^2 L^6\, \frac{4\pi r^2 \delta r}{L^3}, \label{delta_L}
\end{equation}where $U_{t}$ is the typical size of the net translational velocity of a structure. Assuming any two shells are uncorrelated, the total square angular momentum is simply the sum over all shells of Eq.~\eqref{delta_L}. Integrating over $r$, we get
\begin{equation}
    \langle |\bL_V|^2 \rangle \sim U_{t}^2 L^3 R^5. \label{L2_R5}
\end{equation}Like Eq.~\eqref{Pscaling}, the scaling~\eqref{L2_R5} is also adjusted by incompressibility when correlations fall off quickly with distance. This is because
\begin{align}
    \bL_V = \int_V \dd ^3 \br \,\br \times \bu = \int_V \dd ^3 \br \,\br \times (\nabla \times \bpsi).
\end{align}Integrating by parts and expanding the double cross product gives
\begin{align}
    \left(\bL_V\right)_i = - \int_{\partial V} \dd S \left(\delta_{ij}-\frac{r_i r_j}{r^2}\right)r \psi_j + 2\int_V \dd^3 \br\, \psi_i, \label{incompressible_L}
\end{align}where we have chosen $V$ to be spherical. Since $\bpsi$ is a random field, the first integral scales as $R^2$, so it dominates over the second, which scales as $R^{3/2}$. Therefore, Eq.~\eqref{incompressible_L} implies
\begin{align}
    \langle |\bL_V|^2 \rangle \sim U_{t}^2 L^4 R^4. \label{L2_R4}
\end{align} As above, whether ultimately Eq.~\eqref{L2_R5} or Eq.~\eqref{L2_R4} provides the correct scaling depends on the strength of long-range correlations between eddies. 
A formal derivation of these statements may be found in \cite{Davidson13}.

In either case, the scaling of $\langle|\bL_V|^2\rangle$ is different to the na\"{i}ve expectation, $\langle|\bL_V|^2\rangle\sim R^3$, which assumes all eddies to have no translational motion. In that case, the angular momentum of an eddy at a distance $r$ from the origin is $\sim \left[(r+L)U_{r}-(r-L)U_{r}\right]L^3 \sim U_{r} L^4$, where $U_r$ is the typical rotational velocity of a structure. Then Eq.~\eqref{delta_L} becomes
\begin{equation}
    \delta \langle |\bL|^2 \rangle \sim U_{r}^2 L^8\, \frac{4\pi r^2 \delta r}{L^3}, \label{delta_L2}
\end{equation}whence
\begin{equation}
    \langle |\bL_V|^2 \rangle \sim U_{r}^2 L^5 R^3. \label{naive_loitsyansky}
\end{equation}


In summary, while the scalings of $\langle |\bP_V|^2 \rangle$ and $\langle |\bL_V|^2 \rangle$ with $R$ are modified from the na\"{i}ve $R^3$, the correct application of the random-walk argument, taking into account incompressibility, and the greater contribution of more distant structures to $\langle |\bL_V|^2 \rangle$, does result in the correct scalings. We therefore do not consider there to be an essential problem with the application of the random-walk argument in our treatment of the Saffman helicity invariant in Section~\ref{sec:formal}, or in our discussion of general invariants in Section~\ref{sec:general_principles}, though it should be understood that some of these scalings may have to be modified for particular conserved quantities that do not scale with $R$ in the na\"{i}ve way.

\section{Numerical setup and simulation details \label{sec:app_plaws}}

In this work, we have presented numerical simulations conducted with the spectral MHD code Snoopy~\cite{Lesur15}. The code solves the equations of incompressible MHD with hyper-viscosity and hyper-resistivity both of order~$n$, viz.,
\begin{align}
    \frac{\p \bu}{\p t}+\bu\bcdot \bnabla \bu & = -\bnabla p + (\bnabla \times \bB)\times \bB + \nu_n \nabla^{n}\bu, \\
    \frac{\p \bB}{\p t} & = \bnabla \times (\bu \times \bB) + \eta_{n}\nabla^n\bB,\label{mhd}
\end{align}where $p$, the thermal pressure, is determined via the incompressibility condition
\begin{equation}
    \bnabla \bcdot \bu =0.
\end{equation}In all our runs, $\mathrm{Pm}\equiv \nu_n / \eta_n = 1$. The code employs a pseudo-spectral algorithm in a periodic box of size $2\pi$, with a $2/3$ dealiasing rule. Snoopy performs time integration of non-dissipative terms using a low-storage, third-order, Runge-Kutta scheme, whereas dissipative terms are solved using an implicit method that preserves the overall third-order accuracy of the numerical scheme (comparisons between the results of Snoopy and other popular MHD codes for various nonlinear problems may be found in~\cite{Fromang07, Gong20, Squire20}; also see \cite{KunzLesur13} for a test of Snoopy's Hall-MHD module). Units of time are chosen so that the initial magnetic energy is $E_M = 1/2$ (i.e., the unit of time is the initial \Alfven~crossing time of the box). 

We initialise the simulations with a magnetic field whose Fourier representation is
\begin{equation}
    B_i(\bk) = \left[i \epsilon_{ijk}\frac{k_k}{k}+s P_{ij}(\bk)\right]F(k)\,G_j(\bk), \label{initial_field}
\end{equation}where $P_{ij}(\bk) = \delta_{ij}-k_i k_j/k^2$ is the projection operator perpendicular to $\bk$, $G_i(\bk)$ is the Fourier transform of a two- or three-dimensional Gaussian random field with zero correlation length. The parameter $s$ controls the helicity --- $s = 1$ for a helical field, $s = 0$ for a non-helical field --- and is related to the fractional helicity, $\sigma$, discussed in Section~\ref{sec:fractional} by $\sigma = 2s/(1+s^2)$ \cite{Brandenburg19}. The function $F(k)$ sets the initial spectrum of the field:
\begin{equation}
    F(k) = A\begin{cases} k^{(a-D+1)/2}, & k<k_c, \\ 
k^{(a-D+1)/2}\exp(1-k^2/k_c^2), & k>k_c, \end{cases}\label{F_function}
\end{equation}where $D = 2,\,3$ is the number of spatial dimensions, $a$ is the initial spectral exponent of the sub-inertial range (small $k$), and $k_c$ sets the initial peak of the magnetic-energy spectrum, $k_p$, via $k_p = (7/4)^{1/2} \,k_c$. In all runs, we set $a=7$, and $k_c=25 \implies k_p \simeq 33$, so that the magnetic energy is initialised at scales $\simeq 1/33$ of the box size. We note that the initial $k^7$ sub-inertial-range spectrum is different from the $k^4$ spectrum (or $k^3$ in two dimensions) that the system quickly establishes in the subsequent evolution (see Figs.~\ref{fig:helical_spectrum} and~\ref{fig:nonhelical_spectrum}; see also \cite{Davidson13} for a discussion of the role of the large-scale spectrum in decaying hydrodynamic turbulence). Our choice to initialise the simulation with this spectrum was motivated by a finding in our exploratory runs that the transient period before the system entered a period of power-law decay was shorter when the large-scale slope was allowed to establish itself organically. Presumably, this is because even if the spectral exponent is the right one, the structure of the synthetic field, Eq.~\eqref{initial_field}, is not, so it is better not to prejudice the system and let it decide for itself what structures to create at large scales ($k<k_p$).

\begin{figure*}
\includegraphics[width=2\columnwidth]{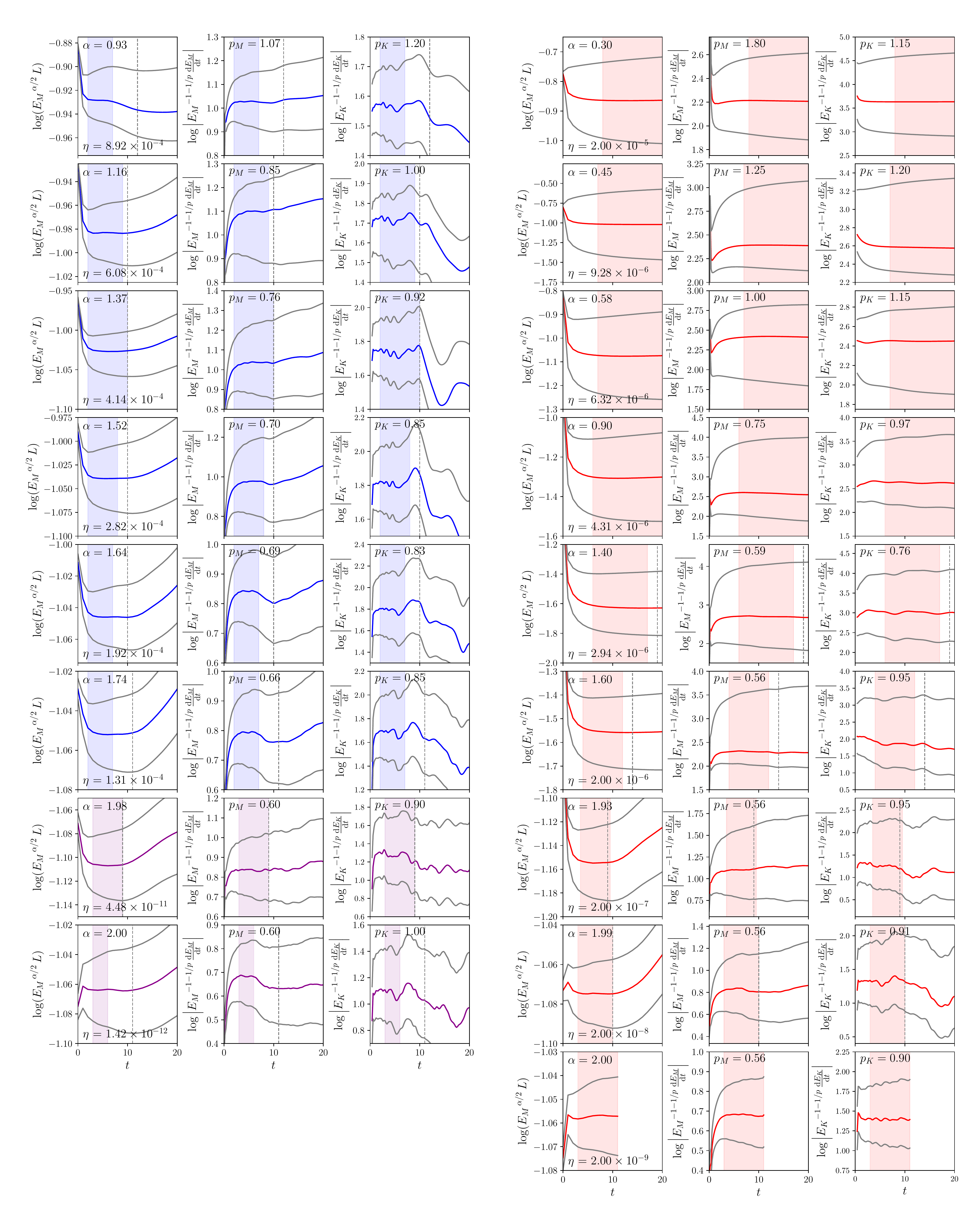}
\caption{\label{fig:helcity_superfigure}Plots used to obtain the parameters $\alpha$ (left), $p_M$ (centre) and $p_K$ (right), for each helical simulation. Simulations with $n=2$, $n=4$, and $n=6$ hyper-dissipation are plotted in blue, red and magenta, respectively. In each case, a horizontal line indicates agreement with the stated value. The shaded region indicates the times at which the decay laws appear to be valid. Grey lines correspond to the values of $\alpha$, $p_M$, and $p_K$ at the extremes of the error bars shown in Figs.~\ref{fig:p_alpha_nonhelical} and \ref{fig:nonhelical_spectrum}. Dashed lines show the time at which $L = 2\pi/5$, i.e., when magnetic structures have scale $1/5$ of the box size, which we find to be roughly the time at which finite-box-size effects begin to affect the decay laws.}
\end{figure*}
\begin{figure*}
\includegraphics[width=2\columnwidth]{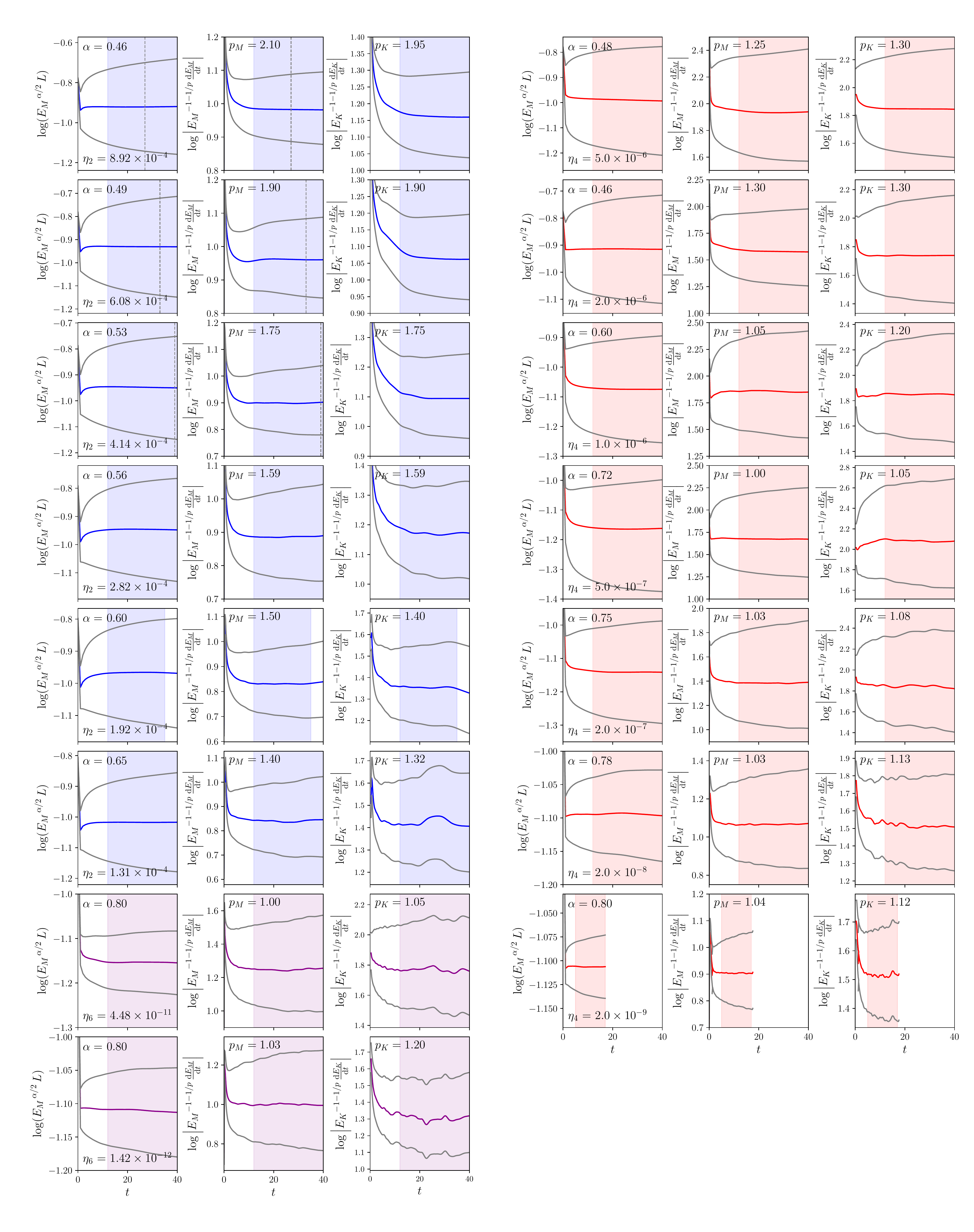}
\caption{\label{fig:nonhelcity_superfigure}Same as Fig.~\ref{fig:helcity_superfigure}, but for non-helical simulations.}
\end{figure*}
\begin{figure*}
\includegraphics[width=2\columnwidth]{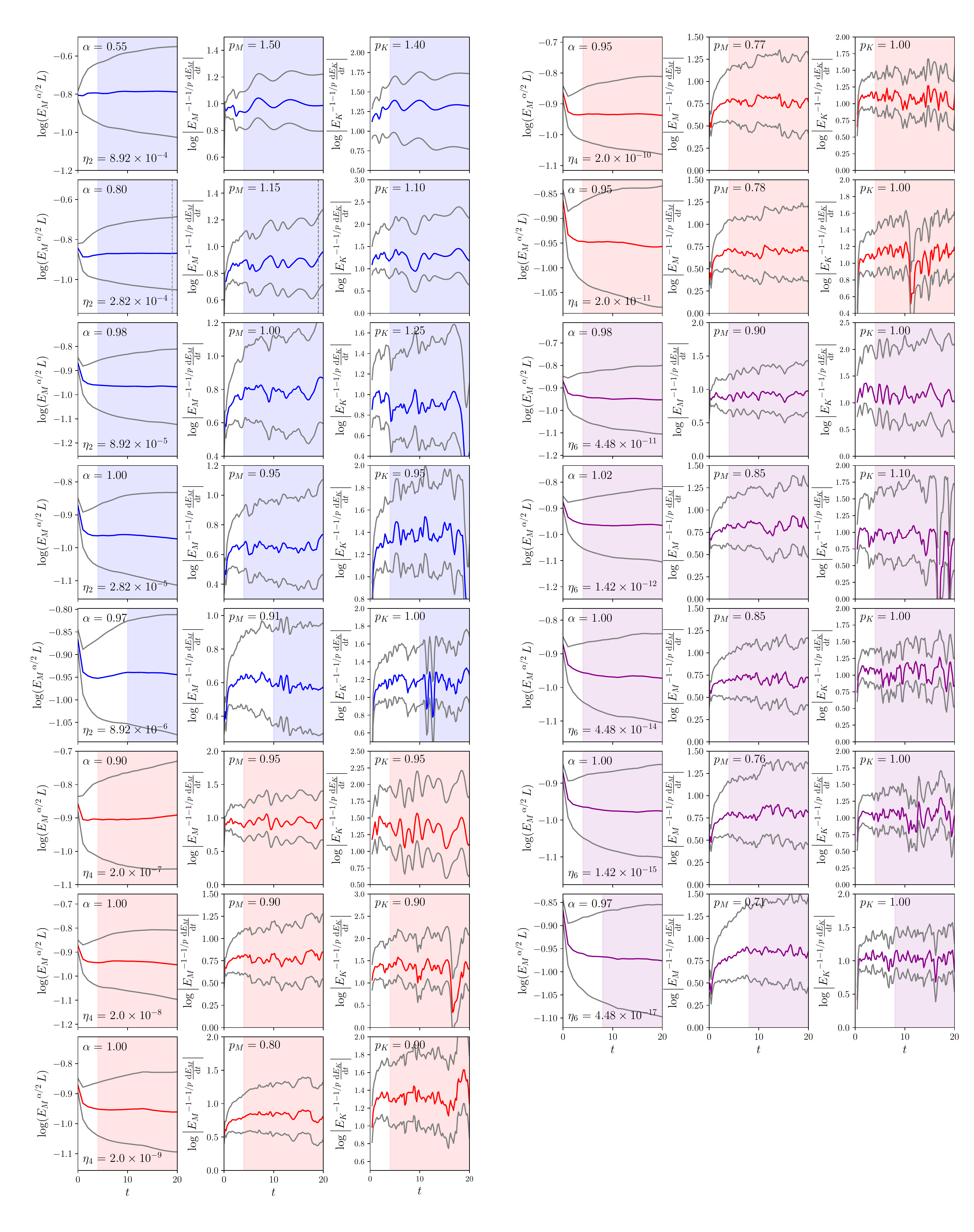}
\caption{\label{fig:2d_superfigure} Same as Fig.~\ref{fig:helcity_superfigure}, but for two-dimensional simulations.}
\end{figure*}

We measure the decay exponents $p_M$ and $p_K$ by plotting $| {E_i}^{1+1/p_i}\dd E_i/\dd t |$
against time, selecting the parameter $p_i$ ($i = M,\,K$) so as to obtain a flat curve. As noted in~\cite{BiskampMuller99}, plots of this type give an unbiased estimate of the decay laws, as compared to more conventional logarithmic plots of $E$ against $t$. The reason for this is that, because the power-law behaviour is only established after a short time~$t_0$ following the initialization of the simulation, a plot of $E\sim (t-t_0)^{-p}$ against $t$ has a bias towards large energies, which decreases over time, giving the false impression of a steeper power law. Furthermore, a logarithmic $t$-axis exaggerates the importance of the initial times, during which the system has, in fact, not established a steady-state decay. In a similar way, we establish the value of $\alpha$ in Eq.~\eqref{BalphaL} by plotting $E_M^{\alpha/2}L$ against time and selecting the value of $\alpha$ to give a flat curve. The power laws and values of $\alpha$ thus obtained are given for all our runs in Table \ref{tab:table}, together with the resolution and initial Lundquist number for each run. For reference, the plots from which these results are obtained are shown in Figs. \ref{fig:helcity_superfigure}, \ref{fig:nonhelcity_superfigure} and \ref{fig:2d_superfigure}. Also plotted there are the curves obtained using the values of $p_M$, $p_K$, and $\alpha$ at the extremes of the error bars in Figs. \ref{fig:p_alpha_helical}, \ref{fig:helical_spectrum}(b), \ref{fig:p_alpha_nonhelical}, \ref{fig:nonhelical_spectrum}(b) and \ref{fig:EK_2d}, to give a sense of the precision with which these results hold.

\renewcommand{\arraystretch}{1.1}
\begin{table*}[]
    \centering
    \begin{tabular}{p{2cm} p{2cm} p{2cm} p{2cm} p{2cm} p{2cm} p{2cm} p{2cm}}
    \hline\hline
         \rule{0pt}{3ex}Type & $n$ & Resolution & $\eta_n = \nu_n$ & $S_{n,\, 0}^{1/n}$ & $\alpha$ & $p_M$ & $p_K$  \\ [0.5ex] 
    \hline
         \rule{0pt}{3ex}H. & 2 & $576^3$ & $8.92 \times 10^{-4}$ & 14.59 & 0.93 & 1.07 & 1.20\\
H. & 2 & $576^3$ & $6.08 \times 10^{-4}$ & 17.68 & 1.16 & 0.85 & 1.00\\
H. & 2 & $576^3$ & $4.14 \times 10^{-4}$ & 21.42 & 1.37 & 0.76 & 0.92\\
H. & 2 & $576^3$ & $2.82 \times 10^{-4}$ & 25.93 & 1.52 & 0.70 & 0.85\\
H. & 2 & $576^3$ & $1.92 \times 10^{-4}$ & 31.44 & 1.64 & 0.69 & 0.83\\
H. & 2 & $576^3$ & $1.31 \times 10^{-4}$ & 38.09 & 1.74 & 0.66 & 0.85\\
[0.5ex]\hline\rule{0pt}{3ex}H. & 4 & $576^3$ & $2.00 \times 10^{-5}$ & 4.30 & 0.30 & 1.80 & 1.15\\
H. & 4 & $576^3$ & $9.28 \times 10^{-6}$ & 5.21 & 0.45 & 1.25 & 1.20\\
H. & 4 & $576^3$ & $6.32 \times 10^{-6}$ & 5.74 & 0.58 & 1.00 & 1.15\\
H. & 4 & $576^3$ & $4.31 \times 10^{-6}$ & 6.32 & 0.90 & 0.75 & 0.97\\
H. & 4 & $576^3$ & $2.94 \times 10^{-6}$ & 6.95 & 1.40 & 0.59 & 0.76\\
H. & 4 & $576^3$ & $2.00 \times 10^{-6}$ & 7.65 & 1.60 & 0.56 & 0.95\\
H. & 4 & $576^3$ & $2.00 \times 10^{-7}$ & 13.61 & 1.93 & 0.56 & 0.95\\
H. & 4 & $576^3$ & $2.00 \times 10^{-8}$ & 24.20 & 1.99 & 0.56 & 0.91\\
H. & 4 & $1152^3$ & $2.00 \times 10^{-9}$ & 43.03 & 2.00 & 0.56 & 0.90\\
[0.5ex]\hline\rule{0pt}{3ex}H. & 6 & $576^3$ & $4.48 \times 10^{-11}$ & 13.29 & 1.98 & 0.60 & 0.90\\
H. & 6 & $576^3$ & $1.42 \times 10^{-12}$ & 23.64 & 2.00 & 0.60 & 1.00\\
[0.5ex]\hline\rule{0pt}{3ex}NH. & 2 & $576^3$ & $8.92 \times 10^{-4}$ & 14.59 & 0.46 & 2.10 & 1.95\\
NH. & 2 & $576^3$ & $6.08 \times 10^{-4}$ & 17.68 & 0.49 & 1.90 & 1.90\\
NH. & 2 & $576^3$ & $4.14 \times 10^{-4}$ & 21.42 & 0.53 & 1.75 & 1.75\\
NH. & 2 & $576^3$ & $2.82 \times 10^{-4}$ & 25.93 & 0.56 & 1.59 & 1.59\\
NH. & 2 & $576^3$ & $1.92 \times 10^{-4}$ & 31.44 & 0.60 & 1.50 & 1.40\\
NH. & 2 & $576^3$ & $1.31 \times 10^{-4}$ & 38.09 & 0.65 & 1.40 & 1.32\\
[0.5ex]\hline\rule{0pt}{3ex}NH. & 4 & $576^3$ & $5.0 \times 10^{-6}$ & 6.09 & 0.48 & 1.25 & 1.30\\
NH. & 4 & $576^3$ & $2.0 \times 10^{-6}$ & 7.65 & 0.46 & 1.30 & 1.30\\
NH. & 4 & $576^3$ & $1.0 \times 10^{-6}$ & 9.10 & 0.60 & 1.05 & 1.20\\
NH. & 4 & $576^3$ & $5.0 \times 10^{-7}$ & 10.82 & 0.72 & 1.00 & 1.05\\
NH. & 4 & $576^3$ & $2.0 \times 10^{-7}$ & 13.61 & 0.75 & 1.03 & 1.08\\
NH. & 4 & $576^3$ & $2.0 \times 10^{-8}$ & 24.20 & 0.78 & 1.03 & 1.13\\
NH. & 4 & $1152^3$ & $2.0 \times 10^{-9}$ & 43.03 & 0.80 & 1.04 & 1.12\\
[0.5ex]\hline\rule{0pt}{3ex}NH. & 6 & $576^3$ & $4.48 \times 10^{-11}$ & 13.29 & 0.80 & 1.00 & 1.05\\
NH. & 6 & $576^3$ & $1.42 \times 10^{-12}$ & 23.64 & 0.80 & 1.03 & 1.20\\
[0.5ex]\hline\rule{0pt}{3ex}2D & 2 & $1152^2$ & $8.92 \times 10^{-4}$ & 14.59 & 0.55 & 1.50 & 1.40\\
2D & 2 & $1152^2$ & $2.82 \times 10^{-4}$ & 25.93 & 0.80 & 1.15 & 1.10\\
2D & 2 & $2304^2$ & $8.92 \times 10^{-5}$ & 46.15 & 0.98 & 1.00 & 1.25\\
2D & 2 & $2304^2$ & $2.82 \times 10^{-5}$ & 82.01 & 1.00 & 0.95 & 0.95\\
2D & 2 & $4608^2$ & $8.92 \times 10^{-6}$ & 145.94 & 0.97 & 0.91 & 1.00\\
[0.5ex]\hline\rule{0pt}{3ex}2D & 4 & $1152^2$ & $2.0 \times 10^{-7}$ & 13.61 & 0.90 & 0.95 & 0.95\\
2D & 4 & $1152^2$ & $2.0 \times 10^{-8}$ & 24.20 & 1.00 & 0.90 & 0.90\\
2D & 4 & $2304^2$ & $2.0 \times 10^{-9}$ & 43.03 & 1.00 & 0.80 & 0.90\\
2D & 4 & $2304^2$ & $2.0 \times 10^{-10}$ & 76.52 & 0.95 & 0.77 & 1.00\\
2D & 4 & $4608^2$ & $2.0 \times 10^{-11}$ & 136.08 & 0.95 & 0.78 & 1.00\\
[0.5ex]\hline\rule{0pt}{3ex}2D & 6 & $1152^2$ & $4.48 \times 10^{-11}$ & 13.29 & 0.98 & 0.90 & 1.00\\
2D & 6 & $1152^2$ & $1.42 \times 10^{-12}$ & 23.64 & 1.02 & 0.85 & 1.10\\
2D & 6 & $2304^2$ & $4.48 \times 10^{-14}$ & 42.04 & 1.00 & 0.85 & 1.00\\
2D & 6 & $2304^2$ & $1.42 \times 10^{-15}$ & 74.76 & 1.00 & 0.76 & 1.00\\
2D & 6 & $4608^2$ & $4.48 \times 10^{-17}$ & 132.95 & 0.97 & 0.71 & 1.00\\
    \hline
    \end{tabular}
    \caption{Details of all simulations, together with the measured decay exponents and the values of $\alpha$ such that $B^\alpha L\sim \const$.}
    \label{tab:table}
\end{table*}

\clearpage
\bibliography{decay_mod}

\end{document}